\documentclass[prx,twocolumn,floatfix,superscriptaddress]{revtex4}
\usepackage[colorlinks]{hyperref}
\usepackage{graphicx}
\usepackage{dcolumn}
\usepackage{bm}
\usepackage{amsmath}
\usepackage{braket}
\usepackage{siunitx}
\usepackage{hyperref}
\usepackage{multirow}

\def\ket#1{|{#1}\rangle}
\def\bra#1{\langle {#1}|}
\begin{document}

\title{Demonstration of nonstoquastic Hamiltonian in coupled superconducting flux qubits}

\author{I. Ozfidan}
\thanks{These authors contributed equally to this work}
\affiliation{D-Wave Systems Inc., 3033 Beta Avenue, Burnaby BC,
Canada, V5G 4M9}

\author{C. Deng \footnote[3]{Current address: Alibaba Quantum Laboratory, Alibaba Group, Hangzhou, China}}

\thanks{These authors contributed equally to this work}
\affiliation{D-Wave Systems Inc., 3033 Beta Avenue, Burnaby BC,
Canada, V5G 4M9}

\author{A. Y. Smirnov}
\affiliation{D-Wave Systems Inc., 3033 Beta Avenue, Burnaby BC,
Canada, V5G 4M9}

\author{T.~Lanting}
\affiliation{D-Wave Systems Inc., 3033 Beta Avenue, Burnaby BC,
Canada, V5G 4M9}

\author{R.~Harris}
\affiliation{D-Wave Systems Inc., 3033 Beta Avenue, Burnaby BC,
Canada, V5G 4M9}

\author{L.~Swenson}
\affiliation{D-Wave Systems Inc., 3033 Beta Avenue, Burnaby BC,
Canada, V5G 4M9}

\author{J.~Whittaker}
\affiliation{D-Wave Systems Inc., 3033 Beta Avenue, Burnaby BC,
Canada, V5G 4M9}

\author{F.~Altomare}
\affiliation{D-Wave Systems Inc., 3033 Beta Avenue, Burnaby BC,
Canada, V5G 4M9}

\author{M.~Babcock}
\affiliation{D-Wave Systems Inc., 3033 Beta Avenue, Burnaby BC,
Canada, V5G 4M9}

\author{C.~Baron}
\affiliation{D-Wave Systems Inc., 3033 Beta Avenue, Burnaby BC,
Canada, V5G 4M9}

\author{A.J.~Berkley}
\affiliation{D-Wave Systems Inc., 3033 Beta Avenue, Burnaby BC,
Canada, V5G 4M9}

\author{K.~Boothby}
\affiliation{D-Wave Systems Inc., 3033 Beta Avenue, Burnaby BC,
Canada, V5G 4M9}

\author{H.~Christiani}
\affiliation{D-Wave Systems Inc., 3033 Beta Avenue, Burnaby BC,
Canada, V5G 4M9}

\author{P.~Bunyk}
\affiliation{D-Wave Systems Inc., 3033 Beta Avenue, Burnaby BC,
Canada, V5G 4M9}

\author{C.~Enderud}
\affiliation{D-Wave Systems Inc., 3033 Beta Avenue, Burnaby BC,
Canada, V5G 4M9}
\author{B.~Evert}
\affiliation{D-Wave Systems Inc., 3033 Beta Avenue, Burnaby BC,
Canada, V5G 4M9}

\author{M.~Hager}
\affiliation{D-Wave Systems Inc., 3033 Beta Avenue, Burnaby BC,
Canada, V5G 4M9}

\author{A.~Hajda}
\affiliation{D-Wave Systems Inc., 3033 Beta Avenue, Burnaby BC,
Canada, V5G 4M9}

\author{J. Hilton}
\affiliation{D-Wave Systems Inc., 3033 Beta Avenue, Burnaby BC,
Canada, V5G 4M9}

\author{S.~Huang}
\affiliation{D-Wave Systems Inc., 3033 Beta Avenue, Burnaby BC,
Canada, V5G 4M9}
\author{E.~Hoskinson}
\affiliation{D-Wave Systems Inc., 3033 Beta Avenue, Burnaby BC,
Canada, V5G 4M9}
\author{M.W.~Johnson}
\affiliation{D-Wave Systems Inc., 3033 Beta Avenue, Burnaby BC,
Canada, V5G 4M9}
\author{K.~Jooya}
\affiliation{D-Wave Systems Inc., 3033 Beta Avenue, Burnaby BC,
Canada, V5G 4M9}

\author{E.~Ladizinsky}
\affiliation{D-Wave Systems Inc., 3033 Beta Avenue, Burnaby BC,
Canada, V5G 4M9}
\author{N.~Ladizinsky}
\affiliation{D-Wave Systems Inc., 3033 Beta Avenue, Burnaby BC,
Canada, V5G 4M9}

\author{R.~Li}
\affiliation{D-Wave Systems Inc., 3033 Beta Avenue, Burnaby BC,
Canada, V5G 4M9}

\author{A.~MacDonald}
\affiliation{D-Wave Systems Inc., 3033 Beta Avenue, Burnaby BC,
Canada, V5G 4M9}

\author{D.~Marsden}
\affiliation{D-Wave Systems Inc., 3033 Beta Avenue, Burnaby BC,
Canada, V5G 4M9}

\author{G.~Marsden}
\affiliation{D-Wave Systems Inc., 3033 Beta Avenue, Burnaby BC,
Canada, V5G 4M9}

\author{T.~Medina}
\affiliation{D-Wave Systems Inc., 3033 Beta Avenue, Burnaby BC,
Canada, V5G 4M9}
\author{R.~Molavi}
\affiliation{D-Wave Systems Inc., 3033 Beta Avenue, Burnaby BC,
Canada, V5G 4M9}

\author{R.~Neufeld}
\affiliation{D-Wave Systems Inc., 3033 Beta Avenue, Burnaby BC,
Canada, V5G 4M9}
\author{M.~Nissen}
\affiliation{D-Wave Systems Inc., 3033 Beta Avenue, Burnaby BC,
Canada, V5G 4M9}
\author{M.~Norouzpour}
\affiliation{D-Wave Systems Inc., 3033 Beta Avenue, Burnaby BC,
Canada, V5G 4M9}

\author{T.~Oh}
\affiliation{D-Wave Systems Inc., 3033 Beta Avenue, Burnaby BC,
Canada, V5G 4M9}
\author{I.~Pavlov}
\affiliation{D-Wave Systems Inc., 3033 Beta Avenue, Burnaby BC,
Canada, V5G 4M9}
\author{I.~Perminov}
\affiliation{D-Wave Systems Inc., 3033 Beta Avenue, Burnaby BC,
Canada, V5G 4M9}
\author{G.~Poulin-Lamarre}
\affiliation{D-Wave Systems Inc., 3033 Beta Avenue, Burnaby BC,
Canada, V5G 4M9}
\author{M.~Reis}
\affiliation{D-Wave Systems Inc., 3033 Beta Avenue, Burnaby BC,
Canada, V5G 4M9}
\author{T.~Prescott}
\affiliation{D-Wave Systems Inc., 3033 Beta Avenue, Burnaby BC,
Canada, V5G 4M9}
\author{C.~Rich}
\affiliation{D-Wave Systems Inc., 3033 Beta Avenue, Burnaby BC,
Canada, V5G 4M9}

\author{Y.~Sato}
\affiliation{D-Wave Systems Inc., 3033 Beta Avenue, Burnaby BC,
Canada, V5G 4M9}
\author{G.~Sterling}
\affiliation{D-Wave Systems Inc., 3033 Beta Avenue, Burnaby BC,
Canada, V5G 4M9}

\author{N.~Tsai}
\affiliation{D-Wave Systems Inc., 3033 Beta Avenue, Burnaby BC,
Canada, V5G 4M9}
\author{M.~Volkmann}
\affiliation{D-Wave Systems Inc., 3033 Beta Avenue, Burnaby BC,
Canada, V5G 4M9}

\author{W.~Wilkinson}
\affiliation{D-Wave Systems Inc., 3033 Beta Avenue, Burnaby BC,
Canada, V5G 4M9}

\author{J.~Yao}
\affiliation{D-Wave Systems Inc., 3033 Beta Avenue, Burnaby BC,
Canada, V5G 4M9}

\author{M. H. Amin\footnote[2]{corresponding author, e-mail: amin@dwavesys.com}}
\affiliation{D-Wave Systems Inc., 3033 Beta Avenue, Burnaby BC,
Canada, V5G 4M9}
\affiliation{Department of Physics, Simon Fraser University, Burnaby BC,
Canada, V5A 1S6}

\begin{abstract}

Hamiltonian-based quantum computation is a class of quantum algorithms in which the problem is encoded in a Hamiltonian and the evolution is performed by a continuous transformation of the Hamiltonian. Universal adiabatic quantum computing, quantum simulation, and quantum annealing are examples of such algorithms. Up to now, all implementations of this approach have been limited to qubits coupled via a single degree of freedom. This gives rise to a stoquastic Hamiltonian that has no {\em sign problem} in quantum Monte Carlo simulations. In this paper, we report implementation and measurements of two superconducting flux qubits coupled via two canonically conjugate degrees of freedom---charge and flux---to achieve a nonstoquastic Hamiltonian. We perform microwave spectroscopy to extract circuit parameters and show that the charge coupling manifests itself as a $\sigma^y\sigma^y$ interaction in the computational basis.  We observe destructive interference in quantum coherent oscillations between the computational basis states of the two-qubit system. Finally, we show that the extracted Hamiltonian is nonstoquastic over a wide range of parameters.

\end{abstract}

\date{\today}

\maketitle


\section{Introduction}

Early generation processors \cite{Johnson11} for Hamiltonian-based quantum computation were designed to perform quantum annealing (QA),  a heuristic algorithm for finding low-energy configurations of a system \cite{Kadowaki98,Farhi01,Santoro02}. With additional features, other applications such as  quantum simulation \cite{Harris18, King18} and machine learning \cite{Mott17,Amin18} have become possible. Currently available large-scale quantum processors are made from a network of radio-frequency superconducting quantum interference devices (rf-SQUIDs) \cite{Johnson11,Harris09, Harris10}. Interaction between pairs of devices is realized through tunable magnetic coupling of their flux degrees of freedom. The low-energy dynamics of individual rf-SQUIDs are effectively captured with their two lowest-energy eigenstates, allowing one to approximate rf-SQUIDs as qubits, described by Pauli matrices $\sigma^{x,y,z}$. The computational basis states $\vert{\uparrow}\rangle$ and $\vert{\downarrow}\rangle$  (eigenfunctions of $\sigma^z$) correspond to directions of persistent current in the body of the rf-SQUID. This network implements the transverse field Ising model Hamiltonian: 
\begin{equation}\label{H}
    H = -\frac{1}{2}\sum_{i}  \Delta_i \sigma_{i}^x + \sum_{i}  h_i \sigma_{i}^z +  \sum_{i<j}J_{ij}\sigma_{i}^z\sigma_{j}^z,
\end{equation}
where $\Delta_i$ and  $h_i$ are tunneling amplitude and energy bias of qubit $i$, respectively, and $J_{ij}$ is the magnetic coupling strength between qubits $i$ and $j$. Quantum annealing is performed by adjusting $\Delta_i \gg h_i, J_{ij}$ at the beginning of the annealing process and gradually evolving until $\Delta_i \ll h_i, J_{ij}$ at the end. 

One important property of Hamiltonian \eqref{H} is that it is {\em stoquastic}, meaning that its equilibrium properties can be simulated using stochastic algorithms such as quantum Monte Carlo (QMC) \cite{Isakov2016,Amin17,Albash2018}. These algorithms typically operate in a local basis, in which all basis vectors are product states of individual qubits. Otherwise, representation of the basis vectors requires exponential resources. A Hamiltonian is stoquastic if there exists a local basis such that all its off-diagonal elements are real and nonpositive \cite{Bravyi08,Lidar18,Klassen18}. A positive off-diagonal element would cause negative transition probabilities, which cannot be simulated by stochastic processes. This issue is called {\em sign} problem in QMC \cite{Loh1990}. If a Hamiltonian is nonstoquastic, there should exist no local basis in which all off-diagonal elements are negative or zero. For some stoquastic Hamiltonians, finding the local transformation that cures the sign problem is by itself intractable  \cite{Lidar18}. Therefore, proving that a Hamiltonian is nonstoquastic is also intractable. To achieve nonstoquasticity, additional interactions such as  $\sigma_{i}^x\sigma_{j}^x$ or $\sigma_{i}^y\sigma_{j}^y$ are needed in  Hamiltonian \eqref{H}. Such interactions can enhance performance of QA \cite{Hormozi17,Nishimori17}, extend the range of quantum simulations \cite{Love14}, provide a path towards annealing-based universal quantum computation \cite{Aharonov,Biamonte,Jordan}, and enable error supression in quantum annealing \cite{Marvian17,Jiang17}. Nonstoquasticity may also be achieved via nonadiabatic evolution \cite{Vinci}. 

\section{Hamiltonian}

\begin{figure}[h!]
\includegraphics[width = 80 mm]{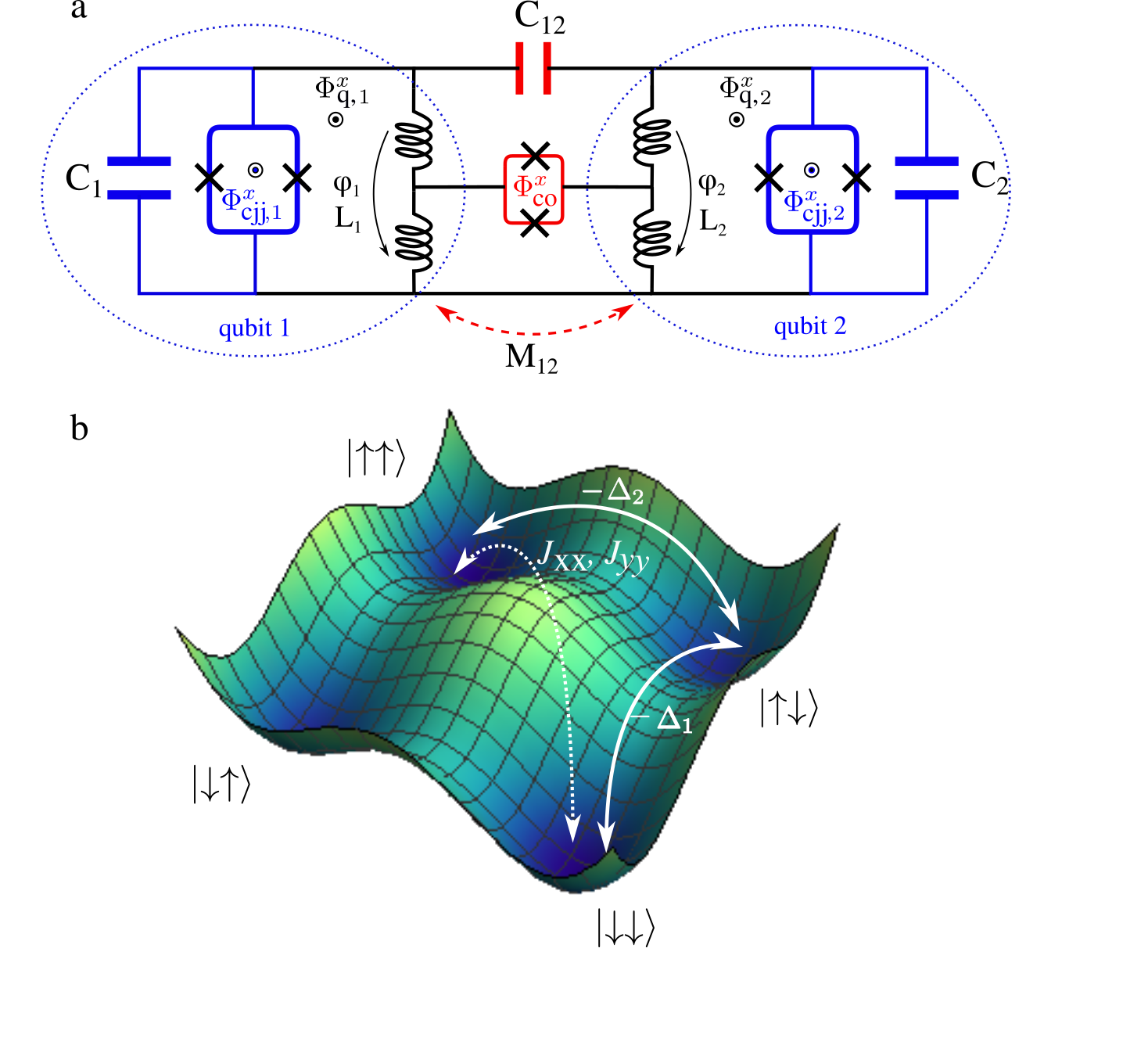}
\caption{Schematic and potential energy of coupled rf-SQUIDs. (a) Schematic circuit of two rf-SQUIDs, coupled inductively via a tunable coupler and capacitively via a fixed capacitor. (b) Effective potential energy of the circuit shown in (a). Arrows show tunneling paths between $\ket{\uparrow\uparrow}$ and  $\ket{\downarrow\downarrow}$ states within the two-state approximation for each rf-SQUID. The solid arrows indicate tunneling due to two consecutive single qubit flips facilitated by $\sigma^x_i$. The dotted arrow represents direct two-qubit cotunneling due to $\sigma^x_1\sigma^x_2$ and $\sigma^y_1\sigma^y_2$. The tunneling amplitudes may have opposite signs, thus leading to destructive interference.} \label{fig:schematics}
\end{figure}

To implement a nonstoquastic Hamiltonian, we use two rf-SQUIDs coupled both inductively, through a tunable mutual inductance $M_{12}$ \cite{Harris09}, and capacitively, through a fixed capacitance $C_{12}$ as shown in Fig.~\ref{fig:schematics}{(a)}. The Hamiltonian of the system is
\begin{eqnarray}
H &=&  \sum_{i=1}^2 \left[ \frac{Q_i^2}{ 2 \widetilde C_i} {+}
\frac{(\Phi_{\text{q},i} {-} \Phi_{\text{q},i}^x)^2}{2 L_i} {-} E_{Ji} (\Phi_{{\rm
cjj},i}^x)  \cos  \frac{ 2 \pi \Phi_{\text{q},i}}{\Phi_0}  \right] \nonumber \\
&&+\frac{M_{12}(\Phi_{\rm co}^x) (\Phi_{\text{q},1} {-} \Phi_{\text{q},1}^x)(\Phi_{\text{q},2} {-} \Phi_{\text{q},2}^x) }{L_1
L_2} \nonumber \\
&& + \frac{C_{12} \, Q_1 Q_2}{C_1 C_2 {+} (C_1 {+} C_2) C_{12} },
\hspace{5mm} \label{HSQUIDs} 
\end{eqnarray}
where $ Q_i$ and $\Phi_{\text{q},i}$ are charge and  flux variables that satisfy the commutation relation $[ \Phi_{\text{q},i}, Q_j] = i \hbar \,\delta_{ij}$, $\Phi_{\text{q},i}^x$ and $\Phi_{{\rm cjj},i}^x$ are external fluxes, and $\Phi_0 = \pi\hbar/e$ is the flux quantum. Each rf-SQUID is characterized by its capacitance $C_i$, inductance $L_i$, and tunable Josephson energy $E_{Ji}(\Phi_{{\rm cjj},i}^x) \approx (\Phi_0/2\pi) I_{ci} \cos (\pi \Phi_{{\rm cjj},i}^x/\Phi_0)$ \cite{Harris10} (for a more detailed description see appendix \ref{rf-SQUID}). The tunable mutual inductance $M_{12}$ is adjusted with the coupler control bias $\Phi_{\rm co}^x$. The renormalized capacitances  are defined as $\widetilde C_{1(2)} = C_{1(2)}+C_{12}C_{2(1)}/(C_{2(1)}+C_{12}) $.  Capacitive couping has been proposed and analyzed theoretically for flux qubits in \cite{Levitov01,Bruder05,Satoh15,Kerman18} and is commonly used in transmon qubits \cite{google,Kounalakis18}.

The potential energy of each rf-SQUID can have a double-well shape 
(see the two-qubit potential in Fig.~\ref{fig:schematics}{(b)}). The barrier height of the potential is controlled by $\Phi_{{\rm cjj},i}^x$, which tunes the tunneling amplitude $\Delta_i$, but also changes the persistent current. The potential is monostable when $\Phi_{{\rm cjj},i}^x =$ 0.5 $\Phi_0$. The flux bias $\Phi_{\text{q},i}^x$ adjusts the tilt of the potential, setting $h_i$. Both $\Phi_{\text{q},i}^x$ and $\Phi_{{\rm cjj},i}^x$ are controlled by high bandwidth coaxial lines, allowing for microwave operation and fast quench of the coherent dynamics. At the end of quench, which involves raising the tunneling barrier rapidly, the persistent current is measured via a shift register coupled to a microwave resonant readout~\cite{Berkley2010,Whittaker16}. 
%

\subsection{Parameter extraction}

Before characterizing the circuit parameters, we calibrate the tunable magnetic coupler, which provides the function $M_{12}(\Phi_{\rm co}^x)$ as discussed in Ref.~\cite{Harris09}. For the rest of the manuscript, we treat the coupler as a simple tunable mutual inductance, assuming dynamics of the coupler are significantly faster than single- and coupled-qubit dynamics. Next, we measure the persistent current of each qubit for a range of $\Phi_{\rm cjj}^x$ in a regime where its tunneling amplitude is negligible and the other qubit is kept monostable. We fit these measurements to a classical rf-SQUID model~\cite{Harris10} and obtain $I_{c,1} = \SI{3.227}{\micro A}$, $L_1 =  \SI{231.9}{\pico H}$, $I_{c,2} = \SI{3.157}{\micro A}$, and $L_2 = \SI{239.0}{\pico H}$.

\begin{figure*}[!ht]
\includegraphics[width = 150 mm]{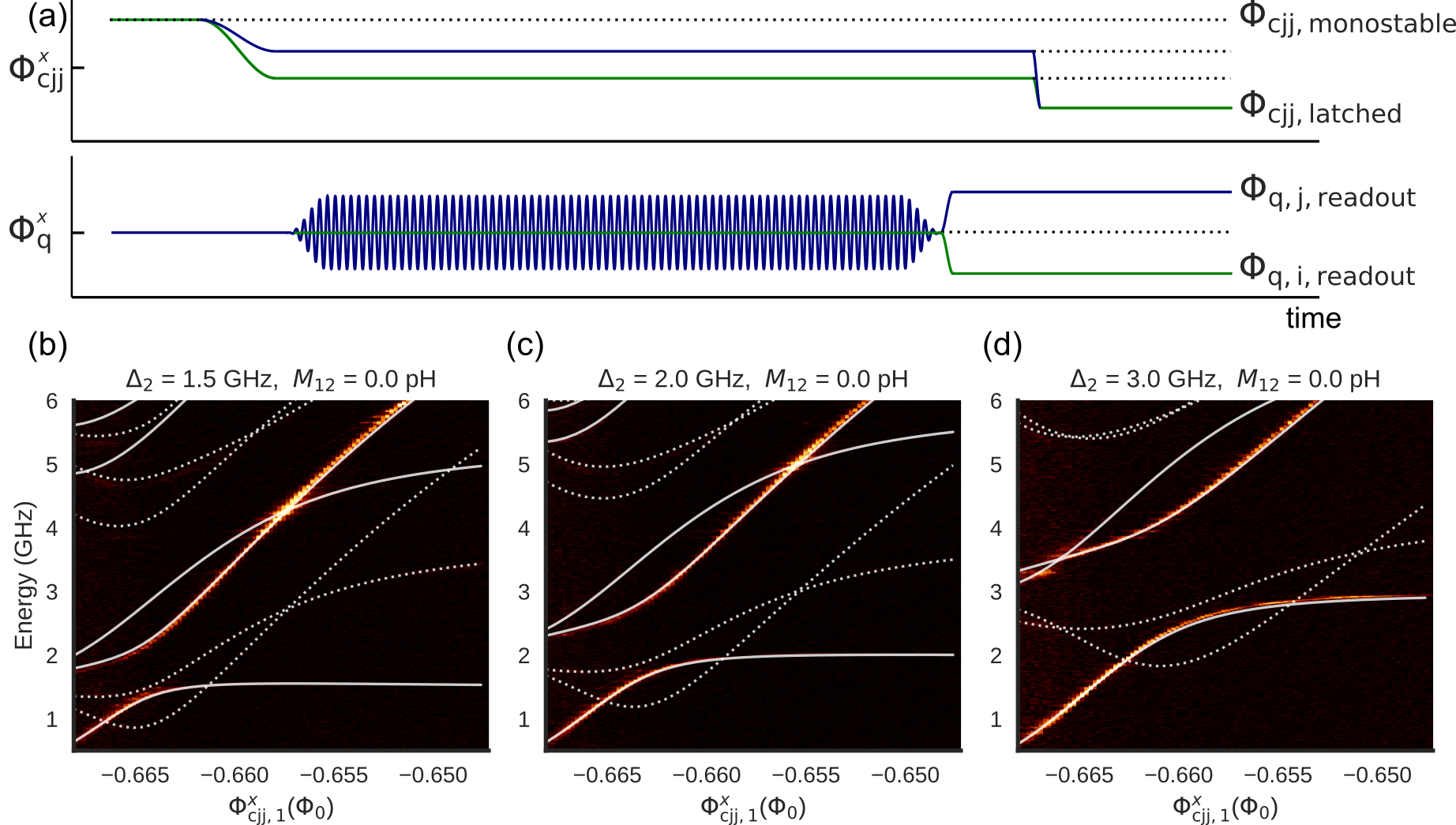}
\caption{Microwave spectroscopy of the coupled two-qubit
system at $M_{12}=0$ and  $\Phi_{q,1}^x = \Phi_{q,2}^x = 0$. (a) Pulse sequences for two-qubit microwave spectroscopy. At $\Phi_{\text{cjj,monostable}}$ the tunneling barrier is at its lowest, resulting in a monostable potential. $\Phi_\text{cjj,latched}^x$ is the opposite, where the tunneling barrier is high and the tunneling amplitude is negligible.  The effective single-qubit spectroscopy has the same pulse sequence except for $\Phi_{{\rm cjj},2}^x=0.5 \Phi_0$. (b-d) Two-qubit spectroscopy versus $\Phi_{{\rm cjj},1}^x$, which controls the qubit 1 barrier height.  Qubit 2 is kept at $\Phi_{{\rm cjj},2}^x$ corresponding to effective single-qubit tunneling amplitudes:  (b) $\Delta_2 = 1.5$ GHz,  (c) $2.0$ GHz, and (d) $3.0$ GHz. Energies are measured relative to the ground state. The solid lines are obtained by fitting the rf-SQUID model, Eq.~(\ref{HSQUIDs}), to the experimental data. Dashed lines represent the excited state energies of uncoupled qubits ($M_{12} =0, C_{12} = 0$) given the other fitting parameters.}
\label{fig:2qspectroscopy}
\end{figure*}

To extract the remainder of the circuit parameters, we perform microwave spectroscopy on the single- and two-qubit systems \cite{Quintana17}. Applying a fixed $\Phi^x_{\rm cjj,2}$ to the second qubit, we sweep the barrier height of the first qubit, controlled by the external flux bias $\Phi^x_{\rm cjj,1}$. At every $\Phi^x_{\rm cjj,1}$, a long microwave pulse ($\SI{1}{\micro s}$) is applied to the first qubit to excite the two-qubit system. The energy eigenstates are read out by applying an adiabatic tilt to both qubits to transform the energy eigenstates into persistent current states, followed by a quench to freeze the dynamics of both qubits before readout. The pulse sequence is shown in Fig.~\ref{fig:2qspectroscopy}{(a)}. The excited state population of each qubit serves as a signal for detecting the energy spectrum of the coupled system. We collect effective single-qubit data by removing the potential barrier of the other qubit, making it monostable, and perform two-qubit spectroscopy for various $\Phi^x_{\rm cjj,2}$. In both sets of experiments we keep $\Phi_{q,1}^x = \Phi_{q,2}^x = 0$,  $M_\text{12} = 0$ pH, while the capacitive coupling is always present. Jointly fitting the effective single- and two-qubit spectroscopy data to the coupled rf-SQUID model, Eq.~(\ref{HSQUIDs}), we obtain the rest of the circuit parameters, $C_{1} = 119.5$~fF, $C_2 = 116.4$~fF, and $C_{12} = 132$~fF. 

Figures \ref{fig:2qspectroscopy}{(b)-(d)} show two-qubit spectroscopy data along with the numerical fit using the rf-SQUID model (\ref{HSQUIDs}). One can see a clear $\Delta_2$ dependent anticrossing that is in good agreement with simulations (solid lines). Without any type of coupling, the spectral lines representing the first excited states of noninteracting qubits would cross as shown by the dashed lines \footnote{\label{capacitivenote}Although the capacitive coupling is off, the capacitive loading on the individual qubits are kept in the simulation.}.  The anticrossing is therefore a signature of capacitive coupling (at $M_{12}=0$) and its energy gap is a measure of the coupling strength. The extracted anticrossing gaps of 0.77, 1.14, and 1.78~GHz at $\Delta_2 = 1.5$, 2.0, and 3.0 GHz, respectively, suggest a strong capacitive coupling.

\subsection{Reduction to qubit model}

We now reduce the continuous rf-SQUID model to a two-state (qubit) model, relevant for quantum computation. The flux degree of freedom is described by $\sigma^z_i$, with $\sigma_1^z\sigma_2^z$ representing the inductive coupling. The charge operator $Q_i {=} - i \hbar \frac{\partial }{\partial \Phi_i}$, on the other hand, is related to $\sigma^y_i$ since both are complex in the computational basis. Thus, the electrostatic coupling between the rf-SQUIDs gives rise to a  $\sigma_1^y\sigma_2^y$ interaction. This term describes direct two-qubit cotunneling. In addition, a  $\sigma_1^x\sigma_2^x$ term is obtained reflecting cotunneling mediated by the high energy states of the rf-SQUIDs.
An effective  two-qubit Hamiltonian can be represented as
\begin{equation} 
H= {-}{\frac{\Delta_1}{ 2}} \sigma_1^x {-}{\frac{\Delta_2 }{2}} \sigma_2^x {+} h_1 \sigma_1^z {+} h_2 \sigma_2^z+ \!\!
\sum_{\alpha,\beta}J_{\alpha\beta} \sigma_{1}^\alpha\sigma_{2}^\beta,
\label{H2state} 
\end{equation}
where $\alpha,\beta = \{x,y,z\}$. Since the continuous Hamiltonian \eqref{HSQUIDs} is real, the reduced Hamiltonian will also be real in the chosen basis, therefore, $J_{\alpha y}=J_{y\alpha}=0$, for $\alpha \ne y$. Parameters of Hamiltonian \eqref{H2state} are  derived from the rf-SQUID model through a reduction techniques described in appendix \ref{Reduction}. Circuit parameters of the rf-SQUID model are extracted by fitting to experimental data obtained from uncoupled ($M_{12}{=}0$) rf-SQUIDs.  


\section{Experimental results}

In this section, we show that the reduced Hamiltonian \eqref{H2state} can explain experimental observations with no further fitting parameters. 

\subsection{Microwave spectroscopy}

\begin{figure*}[t!]
\includegraphics[width = 175 mm]{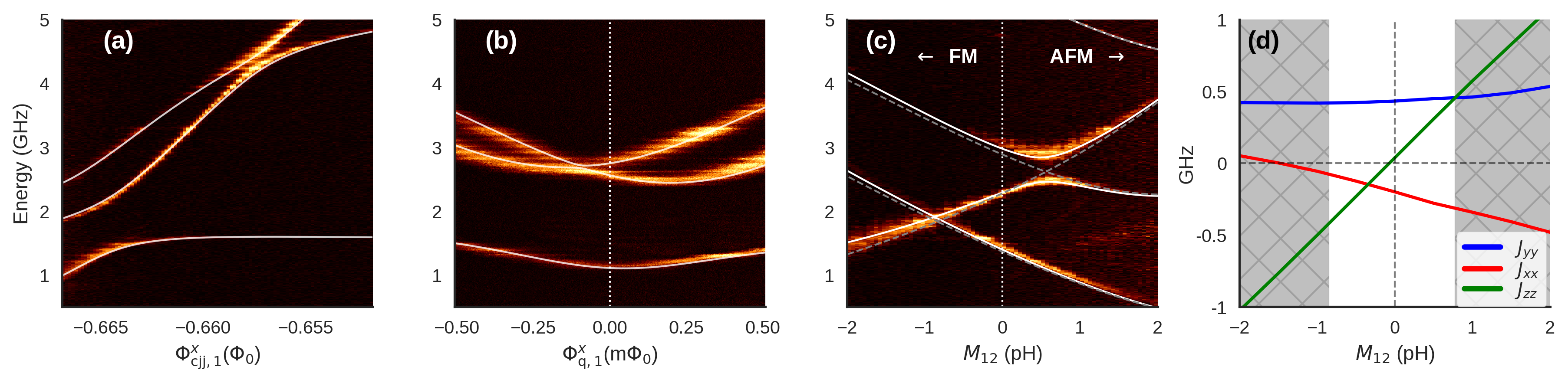}
\caption{Two-qubit spectroscopy at nonzero energy bias $h_i\ne0$. In all panels, qubit 2 is biased away from degeneracy with $\Phi^x_{\rm q,2}=0.1\ {\rm m}\Phi_0$ and $\Phi_{\rm cjj,2}^x$ is set such that the effective single-qubit tunneling is $\Delta_2 = 1.5$ GHz. Energy is measured relative to the ground state.  Solid white lines are numerical simulations obtained using the two-qubit Hamiltonian (\ref{H2state}) with no fitting parameters. (a) Energy spectrum as a function of $\Phi^x_{\rm
cjj,1}$ at fixed $\Phi^x_{\rm q,i}=0.1\ {\rm m}\Phi_0$ and $M_{12}=0$.
(b) Energy spectrum as a function of $\Phi^x_{\rm q,1}$ at $M_{12}$ = 0.55 pH and effective single-qubit  $\Delta_1 = 1.5$ GHz. The vertical dotted line goes through the $\Phi_{q,1}^x=0$ to highlight the asymmetry in the spectrum due to fixed $\Phi_{q,2}^x = 0.1\ {\rm m}\Phi_0$.
(c) Energy spectrum as a function of $M_{12}$ at $\Phi^x_{\rm q,i}=0.1\ {\rm m}\Phi_0$ and $\Delta_i = 1.5$ GHz. The dashed lines correspond to numerical simulations at zero energy biases. The observed avoided crossing is at $M_{12}=0.55$ pH is due to nonzero biases. The vertical dotted line separates ferromagnetic (FM) and antiferromagnetic (AFM) regions. The  asymmetry in the data about this line is due to capacitive coupling. (d) Extracted interaction parameters in Hamiltonian (\ref{H2state}) that provide the theoretical (solid) lines in panel (c) The Hamiltonian is nonstoquastic in the white unshaded area according to Ref.~\cite{Klassen18}. }
\label{fig:spectroscopy}
\end{figure*}

Figures \ref{fig:spectroscopy}{(a)-(c)} depict two-qubit spectroscopy at nonzero energy bias ($h_i\ne0$). Solid white lines in these panels correspond to numerical simulations obtained using Hamiltonian (\ref{H2state}) with no fitting parameters. The presence of a nonzero longitudinal field is necessary for nonstoquasticity, as there exists a unitary transformation that can remove positive off-diagonal elements at $h_i=0$ (see appendix \ref{UTrans}). The experimental parameters in  Fig.~\ref{fig:spectroscopy}{(a)} are the same as those in  Fig.~\ref{fig:2qspectroscopy}{(b)}, except for the flux bias $\Phi_{q,i}^x$ determining $h_i$. This nonzero bias manifests itself in Fig.~\ref{fig:spectroscopy}{(a)} as an avoided level crossing between the second and third excited states. Figure \ref{fig:spectroscopy}{(b)} shows the energy spectrum as a function of $\Phi_{q,1}^x$ while $\Phi_{q,2}^x=0.1$ m$\Phi_0$ and  $M_{12}=0.55$ pH. The top two energy levels cross when  $\Phi_{q,1}^x=-\Phi_{q,2}^x$. Energy spectrum as a function of $M_{12}$ is presented in Fig.~\ref{fig:spectroscopy}{(c)}. As in Fig.~\ref{fig:spectroscopy}{(a)}, the avoided level crossing observed at $M_{12}=0.55$ pH is a result of nonzero energy bias ($h_i\approx 0.15$ GHz at this point). Zero bias simulations are shown by the dashed lines. One can clearly see that the capacitive coupling introduces an asymmetry between the AFM and FM sides of the magnetic coupling. Without the coupling capacitor, the energy spectrum is expected to be symmetric about $M_{12} = 0$.  The theoretical (solid) lines in Fig.~\ref{fig:spectroscopy}{(c)} are produced using the reduced Hamiltonian  \eqref{H2state} with coupling parameters shown in Fig.~\ref{fig:spectroscopy}{(d)}. In  Fig.~\ref{fig:spectroscopy}{(a)-(c)}, theory agrees well with experiment, with no additional fitting.


\subsection{Coherent Oscillations} 

Finally, we measure quantum coherent oscillations in the coupled two-qubit system. Qubits are initialized in a computational basis state by applying a strong flux bias $\Phi_{q,i}^x$. The coherent oscillations are induced by quickly (within 200 ps) pulsing down the barriers of both qubits simultaneously. Before lowering the barriers, the flux bias $\Phi_{q,i}^x$ on each qubit is changed from its value at preparation pulse to its final value. Since the computational basis states are not the eigenstates of the total Hamiltonian, the system undergoes coherent oscillations between these states. After some dwell time $\tau$, the qubits are simultaneously quenched by rapidly raising their energy barriers via $\Phi^x_{\rm cjj,i}$, followed by qubit state readout. We repeat this process for a range of dwell times $\tau$ and coupling strengths $M_{12}$.

\begin{figure*}[!ht]
\includegraphics[width = 160 mm]{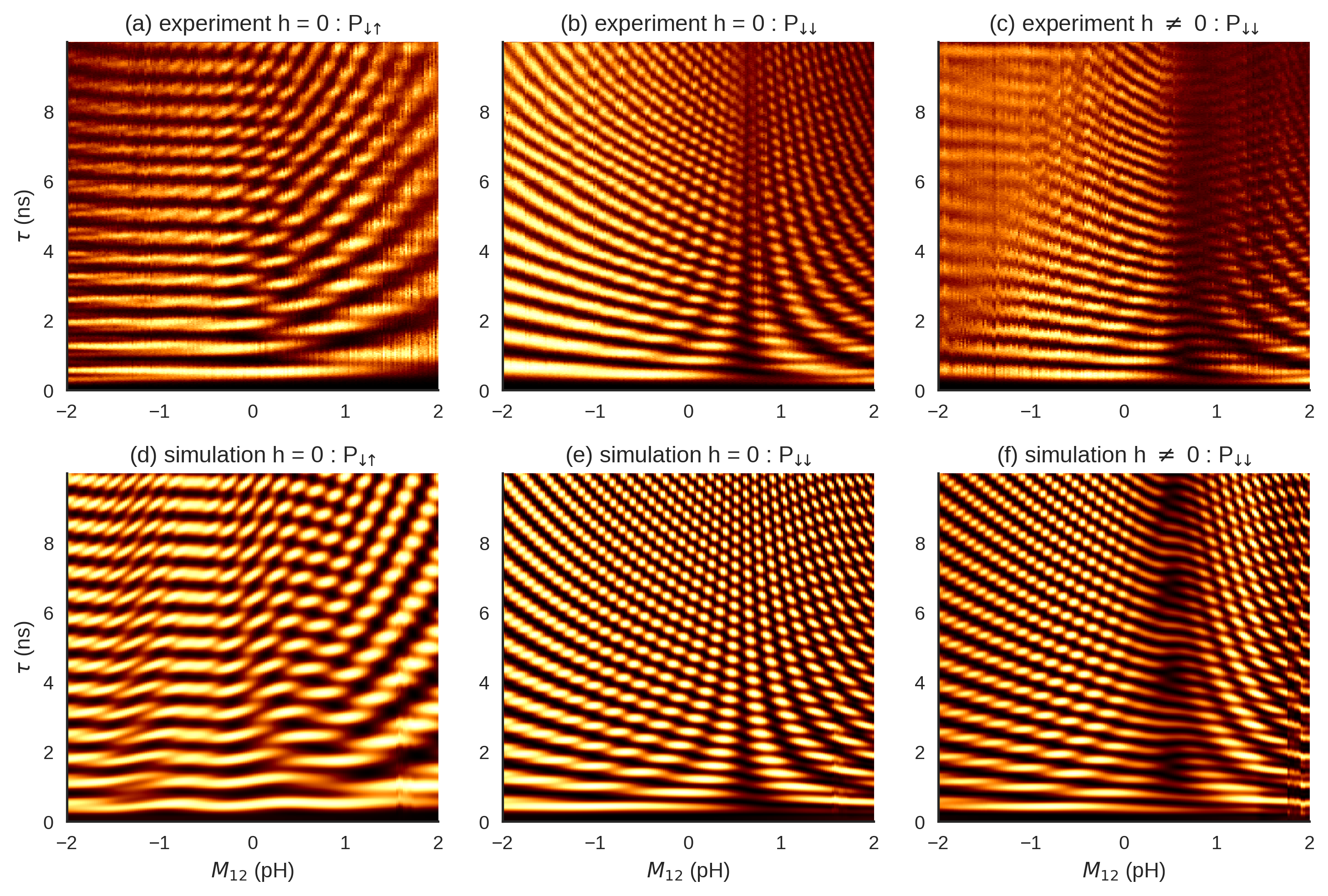}
\caption{Two-qubit coherent oscillations as a function of mutual inductance. (a)
Population of state $\ket{\downarrow \uparrow}$ when the system is
initialized in $\ket{\uparrow\downarrow}$. (b) and (c) Population of
state $\ket{\downarrow \downarrow}$ when the system is initialized
in $\ket{\uparrow\uparrow}$. The final flux-bias on qubits  for panel (c) are set to $0.1$ m$\Phi_0$ corresponding to Figs.~\ref{fig:spectroscopy}{c,d}. Panels (a) and (b) both have a final flux bias of zero. Panels (d), (e), and (f) are numerical simulations obtained using the reduced 4-level model given in Eq.~(\ref{H2state}) corresponding to (a), (b), and (c), respectively.}
\label{fig:coherent-oscillation}
\end{figure*}

In Figs.~\ref{fig:coherent-oscillation}{(a),(b)}, we show the measured state population $P_{\uparrow\downarrow}$ and $P_{\downarrow\downarrow}$ for qubits initially prepared in $\ket{\downarrow\uparrow}$ and $\ket{\uparrow\uparrow}$ configurations, respectively, for  $h_i = 0$. The agreement between time domain experiments and numerical simulations, shown in Fig.~\ref{fig:coherent-oscillation}{(d),(e)}, justifies the two-qubit Hamiltonian (\ref{H2state}) as a valid description of the circuit in Fig.~\ref{fig:schematics}(a). Deviation between theory and experiment can be attributed to decoherence, which is absent in simulations. Single qubit measurements reveal relaxation and dephasing times of $T_1=17$~ns and $T_2=16$~ns, respectively. These timescales are consistent with measurements of isolated flux qubits manufactured in the same stack but without the capacitive coupling.

Features of the energy spectrum as a function of $M_{12}$ (Fig.~\ref{fig:spectroscopy}{(c)}) are reflected in coherent oscillations (Figs.~\ref{fig:coherent-oscillation}{(c),(f)}). The initial configuration, $\ket{\uparrow\uparrow}$,  has significant overlap with the second and third excited states in the AFM region ($M_{12}>0$). The slow oscillation frequency on the right half of figures is therefore related to the gap between these two states. The minimum gap at $M_{12} \approx$ 0.55 pH in Fig.~\ref{fig:spectroscopy}{(c)} corresponds to the maximum slowdown at the same point in Figs.~\ref{fig:coherent-oscillation}{c,f}. When $h_1 = h_2=0$ this gap vanishes (see dashed lines in Fig.~\ref{fig:spectroscopy}{(c)}), nullifying oscillations as seen in Fig.~\ref{fig:coherent-oscillation}{(b)} and \ref{fig:coherent-oscillation}{(e)}. This is a result of the destructive interference between the two-qubit cotunneling channels due to $\sigma_1^x\sigma_2^x$ and $\sigma_1^y\sigma_2^y$ terms and the indirect tunneling channels through sequential single qubit flips caused by  $\sigma_1^x$ and $\sigma_2^x$ terms in Hamiltonian (\ref{H2state}), as schematically illustrated in  Fig.~\ref{fig:schematics}{(b)}. Moreover, we see an additional signature of $J_{yy}$, since without such coupling Fig.~\ref{fig:coherent-oscillation}{(a)} would be a reflection of Fig.~\ref{fig:coherent-oscillation}{(b)} with respect to $M_{12}=0$ (see appendix \ref{CohOsc}).

\section{Stoquasticity} 

While existence of  $J_{xx}$ or  $J_{yy}$ coupling is a necessary condition for nonstoquasticity, it is not sufficient. One needs to show that the sign problem survives all local transformations. While this is intractable at large scales \cite{Lidar18}, it is feasible for two qubits \cite{Klassen18}. To demonstrate this, we extract coefficients in Hamiltonian (\ref{H2state}) which describes Fig.~\ref{fig:spectroscopy}{(c)}. Figure~\ref{fig:spectroscopy}{(d)} plots interaction parameters except for $J_{xz}$ and $J_{zx}$, which are negligibly small. Other parameters are provided in appendix \ref{Reduction}. We see that the electrostatic coupling between the rf-SQUIDs gives rise to a pronounced $J_{yy}$ that is almost constant over the whole range of $M_{12}$. The two-qubit cotunneling mediated by the higher energy states of the rf-SQUIDs leads to a $\sigma_1^x\sigma_2^x$ coupling with a coefficient $J_{xx}$. Both $J_{xx}$ and $J_{zz}$ depend on $M_{12}$ and therefore cannot be tuned independently. The magnitude of $J_{xx}$ is comparable to $J_{yy}$ for large antiferromagnetic coupling. Considering rotations in $x$-$z$ plane, both $J_{xx}$ and $J_{zz}$ terms can reduce the nonstoquastic contribution of $J_{yy}$. The Hamiltonian becomes stoquastic if either of them exceeds $J_{yy}$ in magnitude, as highlighted by the shaded area in the figure. Applying all possible local unitary transformations outlined in Ref.~\cite{Klassen18}, we confirm nonstoquasticity in the unshaded region. Note that $\Delta$, $J_{xx}$ and $J_{yy}$ depend on the barrier heights, with $\Delta$ and $J_{yy}$ being proportional to each other (see appendix \ref{UTrans}). As a result, the width of the nonstoquastic region changes with $\Phi_{{\rm cjj},i}^x$, which controls the qubit tunneling amplitudes $\Delta$. 

\section{Conclusion} 

In summary, we have fabricated two superconducting flux qubits coupled both inductively and capacitively. Starting from an rf-SQUID model, with experimentally extracted circuit parameters, we have obtained a reduced two-qubit Hamiltonian with  $\sigma_1^x\sigma_2^x$, $\sigma_1^y\sigma_2^y$, and  $\sigma_1^z\sigma_2^z$ interactions. We show that the reduced Hamiltonian can explain spectroscopy and coherent oscillation experiments. Considering all local transformations, we prove that the Hamiltonian is nonstoquastic in a wide range of parameters. Equilibrium statistics of such nonstoquastic quantum processors cannot be simulated by QMC due to the sign problem. Implementation of couplings via conjugate degreed of freedom such as charge and flux is an important step towards the development of universal quantum annealers \cite{Aharonov,Biamonte,Jordan}.  Our implementation is based on current, scalable superconducting fabrication technology that is ready to be expanded to a large number of qubits. 
Finally, we should mention that for large scale systems having $\sigma^y_i\sigma^y_j$ terms can make QMC simulation intractable even if the Hamiltonian is stoquastic. This is because finding a local transformation to cure sign problem at large scale is by itself intractable  \cite{Lidar18}.

\section*{Acknowledgements}

We are grateful to D.~Lidar and M.~Marvian for pointing out unitary transformations that cure sign problem at zero bias. We also thank J.~Biamonte, D.~DiVincenzo, I.~Hen, J.~Klassen, P.~Love, J.~Raymond, P.~Saint-Jean, and B.~Terhal for fruitful discussions, and F.~Hanington and A.~King for carefully reading the manuscript.

\appendix

\section{System Hamiltonian}
\label{Hamiltonian}

Our goal here is to derive a Hamiltonian of two rf-SQUIDs that have both, inductive and capacitive, couplings.

\subsection{rf-SQUID with a cjj-loop}
\label{rf-SQUID}

We begin with a description of a single rf-SQUID having a symmetric compound Josephson junction (CJJ) loop, as it is outlined in details in Ref.~\cite{Harris10} (see Fig. 1 in the main text).
The Hamiltonian of this system  has three components:
\begin{equation}
 \label{Hs1} H = H_q + H_{\rm cjj}  - E_J\, \cos \left( \frac{\pi \Phi_{\rm cjj}}{\Phi_0} \right)\, \cos \left( 2 \pi \frac{\Phi_q}{\Phi_0}\right). \end{equation}
Here the Hamiltonians $ H_q$ and $H_{\rm cjj}$ describe the main body of the qubit and the cjj-loop, respectively. They are expressed as
\begin{equation} 
\label{Hs2} 
H_\kappa =  \frac{Q_\kappa^2}{2 C_\kappa}  +
\frac{(\Phi_\kappa - \Phi^x_\kappa)^2}{2 L_\kappa}, \qquad \kappa = \{q, {\rm cjj}\}. 
\end{equation}
The third term in Eq.~(\ref{Hs1}) introduces an interaction between these two loops.
The fluxes in the main body of the qubit and in the cjj-loop are denoted as $\Phi_q$ and $\Phi_{\rm cjj}$, their inductances are  $L_q$ and $L_{\rm cjj}$, respectively. The external fluxes applied to the main body of the qubit and to the cjj-loop are  $\Phi^x_q$ and $\Phi^x_{\rm cjj}.$
The cjj-loop has two symmetric branches $a$ and $b$. Each of them has one Josephson junction, 
with capacitance $C_a (C_b)$ and with a persistent current $I_a (I_b)$.
The total current flowing through both branches of the cjj-loop is denoted as $I_c= I_a+I_b$, with the total Josephson energy $E_J = \frac{\Phi_0 I_c}{2\pi},$ where $\Phi_0 = \pi \hbar/e$ is the flux quantum, $e$ is the electron charge. 
Effective capacitances of the main qubit and the cjj-loop are $C_q = C_a + C_b$ and $C_{\rm cjj} = C_a C_b/(C_a + C_b),$ respectively.
The charge, $Q_\kappa = - i \hbar \frac{\partial}{\partial \Phi_\kappa}$, and flux, $\Phi_\kappa$,
are conjugate operators obeying the commutator: 
$ [ \Phi_\kappa, Q_\kappa]_{-} = i \hbar. $

For qubits 1 and 2 studied in the main text of the paper we have 
$L_{\rm cjj,1} = 17.0$ pH and $L_{\rm cjj,2} = 17.2$ pH, while the body inductances  are $L_1 = 231.9$ pH and $L_2 = 239.0$ pH. When $L_{\rm cjj} \ll L_q,$ the dynamics of the fast degrees of freedom described by the operator $\Phi_{\rm cjj}$ is determined by the ground state of the Hamiltonian $H_{\rm cjj}$  (see  Ref.~\cite{Harris10}). This ground state adiabatically follows the flux degrees of freedom, $\Phi_q,$ in the main body of the rf-SQUID. One can then neglect the kinetic part of the Hamiltonian $H_{\rm cjj}$ and define an effective potential of the main loop by finding the minimum potential energy of the system   (\ref{Hs1}) for each given flux $\Phi_{q}$:
\begin{equation}
 \label{Ueff1}
U_{\rm eff}(\Phi_{q}) = {\rm min}_{\Phi_{\rm cjj}} U(\Phi_{q}, \Phi_{\rm cjj}), \end{equation}
where
\begin{eqnarray}
U(\Phi_{q}, \Phi_{\rm cjj}) = \sum_{\kappa=q,{\rm cjj}} \frac{(\Phi_\kappa - \Phi^x_\kappa)^2}{2 L_\kappa} - \nonumber \\E_J\, \cos \left( \frac{\pi \Phi_{\rm cjj}}{\Phi_0} \right)\, \cos \left( 2 \pi \frac{\Phi_q}{\Phi_0}\right). \end{eqnarray}
This leads to the following effective Hamiltonian for the rf-SQUID,
\begin{equation}
H = \frac{Q_q^2}{2 C_q} + U_{\rm eff}(\Phi_{q}) . \end{equation}
As a rough approximation, the effective potential energy of the rf-SQUID can be written as
\begin{equation} \label{Ueff2} 
U_{\rm eff}(\Phi_{q}) = \frac{(\Phi_q - \Phi^x_q)^2}{2 L_q} - E_J(\Phi_{\rm cjj}^x)\,  \cos \left( 2 \pi \frac{\Phi_q}{\Phi_0}\right), \end{equation}
with a tunable Josephson energy
\begin{equation}
E_J(\Phi_{\rm cjj}^x) = \frac{\Phi_0 I_c}{2\pi} \, \cos \left(\pi \frac{\Phi^x_{\rm cjj}}{\Phi_0}\right). \end{equation}
However, in our simulations, we do not use the above approximation but actually do the minimization with respect to $\Phi_{\rm cjj}$ as given in Eq.~(\ref{Ueff1}) and described in detail in Ref.~\cite{Harris10}. 

We now analyze two rf-SQUIDs connected by a mutual inductance $M_{12}$ and by a capacitor $C_{12}$ as depicted in Fig.~1{a} of the main text.
The inductive coupling between the main loops of the SQUIDs is given by the formula
\begin{equation} \label{UM} 
U_{M} =  \frac{M_{12}}{L_1 L_2} (\Phi_{q,1} - \Phi_{q,1}^x)\,  (\Phi_{q,2} - \Phi_{q,2}^x),
\end{equation}
where $L_i \equiv L_{qi}$ ($i=1,2).$

The kinetic energy of two electrostatically-coupled SQUIDs, with capacitances $C_i \equiv C_{qi},$ has the form
\begin{equation} \label{KinEn}
K = \sum_i \frac{C_i \dot\Phi_{q,i}^2}{2} + \frac{C_{12} (\dot\Phi_{q,2} - \dot\Phi_{q,1} )^2}{2}, \end{equation}
where we use the relation $V_i = \dot \Phi_{q,i} ,$ between a voltage $V_i$ on the $i$-junction and the flux $\Phi_{q,i}.$
Charge $Q_i = \partial {\cal L}/\partial \dot \Phi_{q,i}$ of the $i-$qubit is defined as a derivative of the two-qubit Lagrangian 
\begin{equation} \label{Lagr}
{\cal L} = K - \sum_{i=1}^2 U_{{\rm eff},i}(\Phi_{q,i}) - U_{12},\end{equation}
where $U_{{\rm eff},i}(\Phi_{q,i})$ is the effective potential energy  of the $i-$qubit  (see Eqs.~(\ref{Ueff1}) and (\ref{Ueff2})). Using Eqs.~(\ref{KinEn}, \ref{Lagr}), we obtain the relation between charge and time derivatives of flux,
\begin{eqnarray} 
Q_1 = (C_1 + C_{12}) \dot\Phi_{q,1} - C_{12} \dot\Phi_{q,2}, \nonumber \\
Q_2 = (C_2 + C_{12}) \dot\Phi_{q,2} - C_{12} \dot\Phi_{q,1}.
\end{eqnarray}
We can now write the total Hamiltonian of two coupled rf-SQUIDs, $H = \sum_i \dot\Phi_{q,i} Q_i - {\cal L}$,
 as
\begin{equation} \label{HamT}
H = \sum_{i=1}^2 H_i +  U_M  + U_C. 
\end{equation}
Here 
\begin{equation} \label{HamI}
H_i = \frac{Q_i^2}{ 2 \tilde C_i} {+} U_{{\rm eff},i}(\Phi_{q,i}) \end{equation}
is the Hamiltonian  of $i$-qubit. The qubits  are now characterized by the loaded capacitances: 
\begin{eqnarray} \label{CT}
\tilde C_1 = C_1 + \frac{C_{12} C_2}{C_2 + C_{12} }, \nonumber \\
\tilde C_2 = C_2 + \frac{C_{12} C_1}{C_1 + C_{12} }.
\end{eqnarray}
The inductive interaction between qubits is given by (\ref{UM}).
The capacitive coupling is determined by the potential 
\begin{equation} \label{UC}
U_C =  \frac{C_{12} \, Q_1 Q_2}{C_1 C_2 {+} (C_1
{+} C_2) C_{12} }.
\end{equation}

At large scales, the capacitative network renormalizes the capacitances of each rf-SQUID. In addition, capacitative loading should reduce sensitivity of the qubits to charge fluctuations, thus making this device more immune to charge noise.

\subsection{Reduction approach} 
\label{Reduction}

Diagonalizing the single-SQUID Hamiltonian $H_i$ (\ref{HamI}) we obtain a set of energy eigenstates, $\ket{\chi_{\mu}^i}$, and eigenenergies, $\varepsilon_\mu^i,$ such that 
\begin{equation} \label{Ham1N}
 H_i = \sum_{\mu =1}^{N_i} \varepsilon^i_{\mu} \ket{\chi_{\mu}^i}\bra{\chi_{\mu}^i}. \end{equation}
For each SQUID we take into account a large number, $N_i \gg 1,$ of the energy eigenstates and write the Hamiltonian (\ref{HamT})
 of two coupled rf-SQUIDs in the basis formed by direct products  $ \ket{ \chi_{\mu}^1\otimes \chi_{\nu}^2}  \equiv \ket{\chi_{\mu}^1} \otimes \ket{\chi_{\nu}^2}$,
\begin{eqnarray} \label{Ham2N}
H =  \sum_{\mu =1}^{N_1} \varepsilon^1_{\mu} \ket{\chi_{\mu}^1}\bra{\chi_{\mu}^1}  + 
 \sum_{\nu =1}^{N_2} \varepsilon^2_{\nu} \ket{\chi_{\nu}^2}\bra{\chi_{\mu}^2}  + U_M + U_C. \nonumber \\
\end{eqnarray}
Here charge and flux operators  should be also written in the 
 $\ket{ \chi_{\mu}^1\otimes \chi_{\nu}^2}$ basis. 
The eigenstates, $\ket{\eta_a},$ of the Hamiltonian (\ref{Ham2N}) 
 become
\begin{equation} \label{eta1}
\ket{\eta_a} = \sum_{\mu}^{N_1} \sum_{\nu}^{N_2} c^a_{\mu\nu}  \ket{\chi_{\mu}^1 \otimes \chi_{\nu}^2}, \end{equation}
where the amplitudes are given as
\begin{equation} \label{cMN}
 c^a_{\mu\nu} = \langle \chi_{\mu}^1\otimes \chi_{\nu}^2 | \eta_a   \rangle.\end{equation}
The two-SQUID Hamiltonian,  
\begin{equation} \label{HamD}
 H = \sum_{a=1}^{N} \varepsilon_a \ket{\eta_a}\bra{\eta_a}, \end{equation}
is characterized by the energy spectrum $\varepsilon_a,$ 
with a total number of levels $N = N_1 N_2.$

Working with continuous models becomes computationally challenging beyond a small number of coupled rf-SQUIDs. With the goal to use rf-SQUIDs as qubits, one needs to reduce the continuous Hamiltonian to a discrete (qubit) Hamiltonian. 
For uncoupled SQUIDs, we choose the following superpositions of two lowest-energy states 
with the mixing angle $\theta_i$:
\begin{eqnarray} \label{UpDown}
\ket{\downarrow}_i  &=& \cos \theta_i\, \ket{\chi_1^i } + \sin\theta_i\, \ket{\chi_2^i} \nonumber \\
\ket{\uparrow}_i &=& -\sin \theta_i \, \ket{\chi_1^i} + \cos \theta_i \,\ket{\chi_2^i}.
\end{eqnarray}
The basis states $\ket{\downarrow_i}$ and $\ket{\uparrow_i}$ correspond to the left and right circulating currents, or, equivalently, to the left and right sides of the SQUID potential well. The mixing angle $\theta_i$ is chosen to maximize the left-well population in the state $\ket{\downarrow_i}$ and, thus, the right-well population when the SQUID is in the state $\ket{\uparrow_i}.$
The interaction Hamiltonian mixes the states $\ket{\downarrow_i}, \ket{\uparrow_i}$ with higher energy states of the individual rf-SQUIDs that one needs to take into account for a correct description of the coupled system. 

Since we are interested in the low energy spectrum of the coupled system, the number of eigenstates taken into account in Eqs. (\ref{Ham1N},\ref{Ham2N}) can be truncated to just two states for each rf-SQUID.
With this truncation, four eigenvectors, $\ket{\eta_a}$ ($a = 1, \ldots, 4$),  of the two-SQUID system  are approximated as 
\begin{equation} \label{eta2}
\ket{\eta_a} \simeq \frac{1}{N_a}\,\sum_{\mu=1}^{2} \sum_{\nu=1}^{2} c^a_{\mu\nu}  \ket{\chi_{\mu}^1 \otimes \chi_{\nu}^2}, \end{equation}
where the amplitudes $c^a_{\mu\nu}$ are given in Eq.~(\ref{cMN}),
and the normalization coefficient is calculated as
 $$ N_a = \sqrt{\sum_{\mu=1}^{2} \sum_{\nu=1}^{2} |c^a_{\mu\nu}|^2}.$$ 
We  apply the Gram-Schmidt procedure to the four states given in Eq.~(\ref{eta2}) to obtain the orthonormalized set of the two-qubit basis states obeying the relation: $\langle \eta_a |\eta_b\rangle = \delta_{ab}.$
This reduction approach only works in the limit where $N_a \approx 1$ for $ a=1,\ldots,4$.
In this case the most of the population of two-SQUID system is distributed over the tensor products 
 $\ket{\chi_{\mu}^1 \otimes \chi_{\nu}^2}$ of two lowest eigenstates of the isolated rf-SQUIDs ($\mu,\nu = 1,2$).  

In order to derive the reduced Hamiltonian of two coupled SQUIDs  we start with a Hamiltonian (\ref{HamD}) truncated to the four lowest-energy states. In the energy basis this Hamiltonian is described by the diagonal matrix: $H = {\rm diag} (\varepsilon_1, \varepsilon_2, \varepsilon_3, \varepsilon_4). $ 
The Hamiltonian $H$ can be written in the $\chi-$basis formed by the four vectors $\ket{\chi_{\mu}^1 \otimes \chi_{\nu}^2}$ ($\mu,\nu = 1,2)$ by applying the rotation matrix $R_1 = (\ket{\eta_1}\; \ket{\eta_2}\; \ket{\eta_3}\; \ket{\eta_4})$ such that $H_{\chi} =  R_1 H R_1^T$.
Finally, the unitary matrix $R_2 = (\ket{1}\, \ket{2} \,\ket{3}\, \ket{4}) $  rotates the Hamiltonian into the computational basis, formed by the vectors:
\begin{eqnarray} \ket{1} = \ket{\downarrow_1\otimes \downarrow_2}, \, \ket{2} = \ket{\downarrow_1\otimes \uparrow_2}, \nonumber \\\ket{3} = \ket{\uparrow_1\otimes \downarrow_2}, \, \ket{4} = \ket{\uparrow_1\otimes \uparrow_2}. \end{eqnarray}
 After this rotation the Hamiltonian $ H = R_2 H_{\chi} R_2^T$ can be represented by the 4x4 matrix,
\begin{widetext}
\begin{eqnarray}
 H = \left[\begin{array}{cccc}
J_{zz}-h_1-h_2 &-\Delta_2/2 &-\Delta_1/2 & J_{xx}-J_{yy} \\
-\Delta_2/2&-J_{zz}-h_1+h_2	 & J_{xx}+J_{yy} & -\Delta_1/2 \\
-\Delta_1/2 & J_{xx}+J_{yy} & -J_{zz}+h_1-h_2 &-\Delta_2/2 \\
J_{xx}-J_{yy} &-\Delta_1/2& -\Delta_2/2 &J_{zz}+h_1+h_2 
\end{array}\right],
\label{Hc1}
\end{eqnarray}
\end{widetext}
or by the Hamiltonian of two coupled spins (qubits):
\begin{equation} 
H = {-}{\Delta_1 \over 2} \sigma_1^x {-}{\Delta_2 \over 2} \sigma_2^x {+} h_1 \sigma_1^z {+} h_2 \sigma_2^z+ \!\!
\sum_{\alpha,\beta}J_{\alpha\beta} \sigma_{1}^\alpha\sigma_{2}^\beta,
\label{Hc2} 
\end{equation}
where $\alpha, \beta = {x,y,z}$. We use the following representation of the Pauli matrices:
\begin{eqnarray}
\sigma^x_i = \ket{\downarrow_i}\bra{\uparrow_i} + \ket{\uparrow_i}\bra{\downarrow_i} , \nonumber \\
\sigma^y_i = i (\ket{\downarrow_i}\bra{\uparrow_i} - \ket{\uparrow_i}\bra{\downarrow_i}) , \nonumber \\
\sigma^z_i = \ket{\uparrow_i}\bra{\uparrow_i} - \ket{\downarrow_i}\bra{\downarrow_i}.
\end{eqnarray}
The parameters of the Hamiltonian (\ref{Hc2}) are extracted by comparing the operator  $ H = R_2 H_{\chi} R_2^T$ with the matrix (\ref{Hc1}). 
Note that $J_{yx} = J_{xy} = J_{yz} = J_{yz} = 0,$ since the Hamiltonian is real and the coefficients $J_{xz}$ and $J_{zx}$ are negligible: $J_{xz}, J_{zx} \ll \Delta_1, \Delta_2.$ The non-negligible parameters of the Hamiltonian $H$ describing interactions between the qubits are shown in 
Fig.~\ref{fig:spectroscopy}{(d)} of the main text. 

\begin{figure*}[!]
\includegraphics[width=175mm]{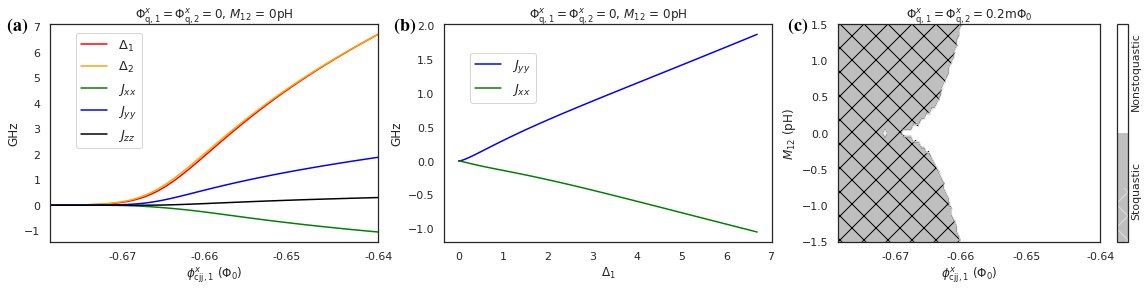}
\caption{(a) Two-qubit Hamiltonian parameters as functions of $\Phi^x_{cjj,i}$, which controls the barrier height, at zero flux bias and zero inductive coupling. (b) $J_{xx}$ and $J_{yy}$ as a function of $\Delta_i$ for the same setting as in (a). (c) Diagram depicting nonstoquastic region as function of mutual inductance and  $\Phi^x_{cjj,i}$ for nonzero flux bias. The Hamiltonian is nonstoquastic in white area.}
\label{fig:JYY-Delta}
\end{figure*}

Figure \ref{fig:JYY-Delta}(a) shows how the Hamiltonian parameters change during the annealing which is performed by changing $\Phi^x_{\rm cjj,i}$. All off-diagonal terms vanish as $\Phi^x_{\rm cjj,i}$ varies from right to left as the annealing proceeds. Linear dependence of $J_{xx}$ and $J_{yy}$ on $\Delta_i$ is evident from Fig.~\ref{fig:JYY-Delta}(b).

One can equivalently use the projection technique of Ref.~\cite{Amin12S} to derive a reduced Hamiltonian for two interacting subsystems, namely our SQUIDs. The reduced Hamiltonian obtained with the projection approach agrees with the two-qubit Hamiltonian  given by Eqs. (\ref{Hc1}) and (\ref{Hc2}).

\subsection{Unitary transformations and stoquasticity}
\label{UTrans}

Hamiltonian (\ref{Hc1}) can have positive off-diagonal matrix elements under several conditions but not all of those make the Hamiltonian nonstoquastic. 
One has to take into account all local transformations of the basis before deciding about nonstoquasticity of the Hamiltonian. Using the unitary transformation $H\to U^\dagger H U$, with $U=\otimes_i(\sigma^z_i)^{(1- {\rm sign}\, \Delta_i)/2}$, one can make all $\Delta_i\sigma^x_i$ terms in the Hamiltonian nonpositive regardless of the sign of $\Delta_i$. Hamiltonian (\ref{Hc1}) has positive matrix elements if $|J_{yy}| > |J_{xx}|$, since matrix elements $H_{14}$ and $H_{23}$ have opposite signs, due to $J_{yy}$, making one of them positive. However, this condition is not sufficient for nonstoquasticity. To demonstrate this fact, we apply a Hadamard rotation on both qubits turning $x-$axis to $z-$axis and vice versa, so that $\sigma^x_i = \tau^z_i, \; \sigma^y_i = \tau^y_i,$ and $\sigma^z_i = - \tau^x_i.$ Here $\tau^x_i, \tau^y_i, \tau^z_i$ are Pauli matrices of $i$-qubit in the rotated frame. 
After the Hadamard rotation the two-qubit Hamiltonian has the form:
\begin{widetext}
\begin{eqnarray}
 H = \left[\begin{array}{cccc}
J_{xx}+\Delta_1/2 +\Delta_2/2 &-h_2&-h_1 & J_{zz}-J_{yy} \\
-h_2  &-J_{xx}+\Delta_1/2-\Delta_2/2	 & J_{zz}+J_{yy} & -h_1 \\
-h_1 & J_{zz}+J_{yy} & -J_{xx}-\Delta_1/2+\Delta_2/2 &-h_2 \\
J_{zz}-J_{yy} &-h_1 & -h_2 &J_{xx}-\Delta_1/2-\Delta_2/2
\end{array}\right].
\label{Hc3}
\end{eqnarray}
\end{widetext}
This Hamiltonian has positive matrix elements at $|J_{yy}| > |J_{zz}|.$ At non-zero biases, $h_i \neq 0,$ (and at non-zero $\Delta_i$) only rotations in the $x$-$z$ plane are allowed \cite{Klassen18} since rotations in  $x$-$y$ and $z$-$y$ planes introduce complex matrix elements in the Hamiltonian. In 
Fig.~\ref{fig:spectroscopy}{(d)} of the main text, we search over all possible local rotations outlined in detail in Ref. \cite{Klassen18} and find, in particular, that at the borders of the nonstoquastic range the $\sigma^z \sigma^z$ coupling is of order of the $\sigma^y \sigma^y$ interaction strength:  $|J_{zz}| \sim |J_{yy}|.$ Figure \ref{fig:JYY-Delta}(c) is an extension to this figure, displaying nonstoquastic region as a function of $M_{12}$ and $\Phi^x_{cjj,i}$, which varies during annealing. It follows from Fig.~\ref{fig:JYY-Delta}(a) and (c) that during the critical region of annealing, when off-diagonal elements of the Hamiltonian are present, the Hamiltonian can be nonstoquastic depending on the magnitude of the coupling. 

The matrix~(\ref{Hc3}) also provides the intuition necessary to see effects of $J_{xx}$ and $h_1, h_2$ terms on experimentally measured spectra. When $J_{zz}=0$ and $h_1 = h_2=0$, the Hamiltonian is block-diagonal as there is no interaction between aligned and anti-aligned states. In the absence of the $J_{xx}$ term, the eigenvalues of the Hamiltonian becomes $ \pm \sqrt{J_{yy}^2 + \left(\frac{\Delta_1\pm\Delta_2}{2}\right)^2}$. Without the $J_{xx}$ term, the two highest eigenvalues can only cross if $\Delta_1 \Delta_2= 0.$ 
However we see a clear level crossing in spectroscopy vs. $\Phi_{\text{cjj},1}^x$ plots in the middle row of Fig.~\ref{fig:2q-spectroscopy-extra} where $\Delta_i \neq 0$. This is a clear signature of a $J_{xx}$ type interaction. Furthermore, the second and third excited states, that cross in presence of $J_{xx}$ belong to two separate blocks of the Hamiltonian. When $h_i\ne0$ the Hamiltonian is no longer block diagonal and the two highest excited states start interacting. Hence, the avoided crossings visible in energy vs $\Phi_{\text{cjj},1}^x$ and $M_{12}$ in 
Fig.~\ref{fig:spectroscopy}{(a),(c)}
of the main text are clear signatures of non-zero $h_{1,2}$.

\section{Qubit and coupler parameter characterization}

\subsection{Coupler characterization}
\label{Coupler}

\begin{figure}[!]
\includegraphics[width=80mm]{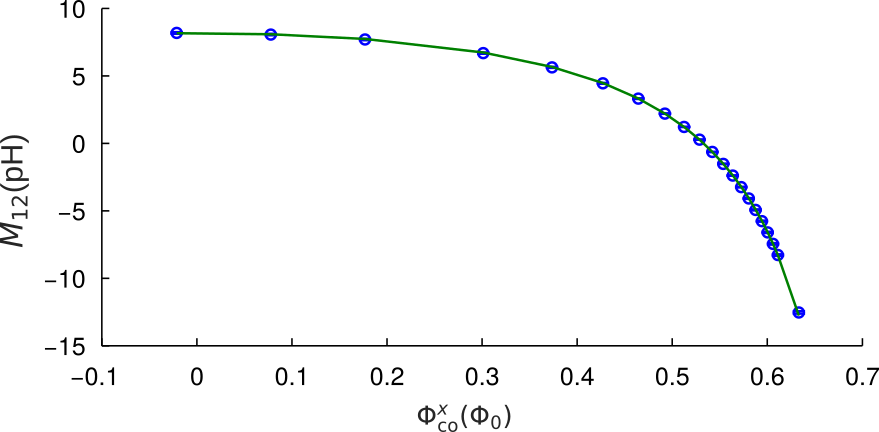}
\caption{Measured Coupler $M_{12}$ vs coupler bias $\Phi_{co}^x$ and a fit to the classical model of the coupler.}
\label{fig:coupler-meff}
\end{figure}

We follow Ref.~\cite{Harris09} for the characterization of the tunable magnetic coupler. In Fig.~\ref{fig:coupler-meff}, we show the measured coupler
$M_{12}$ versus the coupler flux bias $\Phi_{co}^x$. The data is fitted to a classical model:

\begin{equation}
M_{12} = \frac{M_{co,q}^2}{L_{co}}\frac{\beta \cos(\varphi_{co}^x/2)}{1+ \beta \cos (\varphi_{co}^x/2)} + M_{12}^{(0)},
\label{co-meff}
\end{equation}
with $M_{co,q}$ the mutual inductance between the qubit and the coupler, $L_{co}$ the coupler inductance, $\beta = 2\pi L_{co}
I_{c,co}/\Phi_0$, $M_{12}^{(0)}$ the stray mutual inductance between the qubits, and $\varphi_{co}^x = 2\pi \Phi_{co}^x$ the normalized
external bias of the coupler. From the fitting, we obtained $M_{co,q}^2/L_{co} = 10.77$~pH, $\beta = 1.416$, and $M_{eff}^{(0)} = 1.848$~pH.
Using Eq.~(\ref{co-meff}) and the above fitting parameters, we can set the coupler to any coupling strength within the range $|M_{12}| \le 8.145$~pH.

\subsection{Quasi-static qubit characterization}

\begin{figure}[!]
\includegraphics[width=80mm]{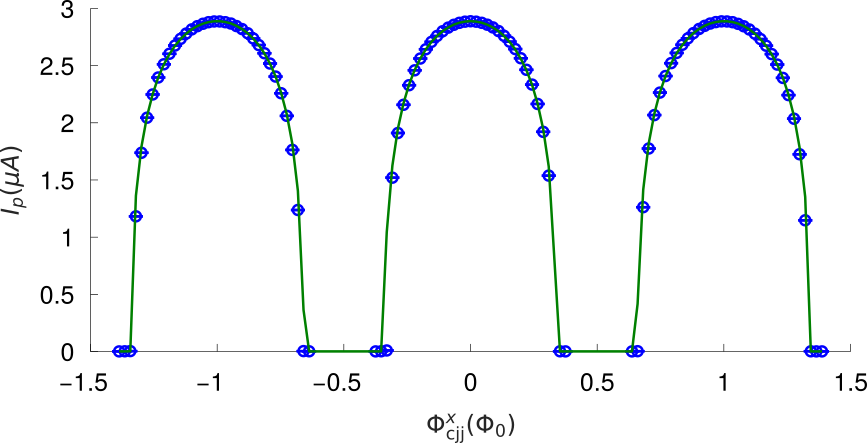}
\caption{Measured persistent current $I_p$ versus $\Phi_{{\rm cjj }}^x$ that controls the barrier height of the double-well potential and a fit to the
classical model of Qubit 1.}
\label{fig:icirc-fit}
\end{figure}

When the tunneling barrier is high, where the single-qubit tunneling is largely suppressed, the properties of a flux qubit can be described by a classical model~\cite{Harris10}. 
In this regime, we measure the qubit persistent current versus the flux bias that control the
barrier height $\Phi_{{\rm cjj}}^x$ across three flux quantum $\Phi_0$ for each qubit individually. Fits to the classical model (see
Fig.~\ref{fig:icirc-fit}) yield the following qubit parameters: $I_{c,1} = 3.227~\mu$A, $I_{c,2} = 3.157~\mu$A, $L_1 = 231.6~$pH, $L_2 = 239.0~$pH.

\subsection{Single-qubit spectroscopy}

\begin{figure}[t]
\includegraphics[width=80mm]{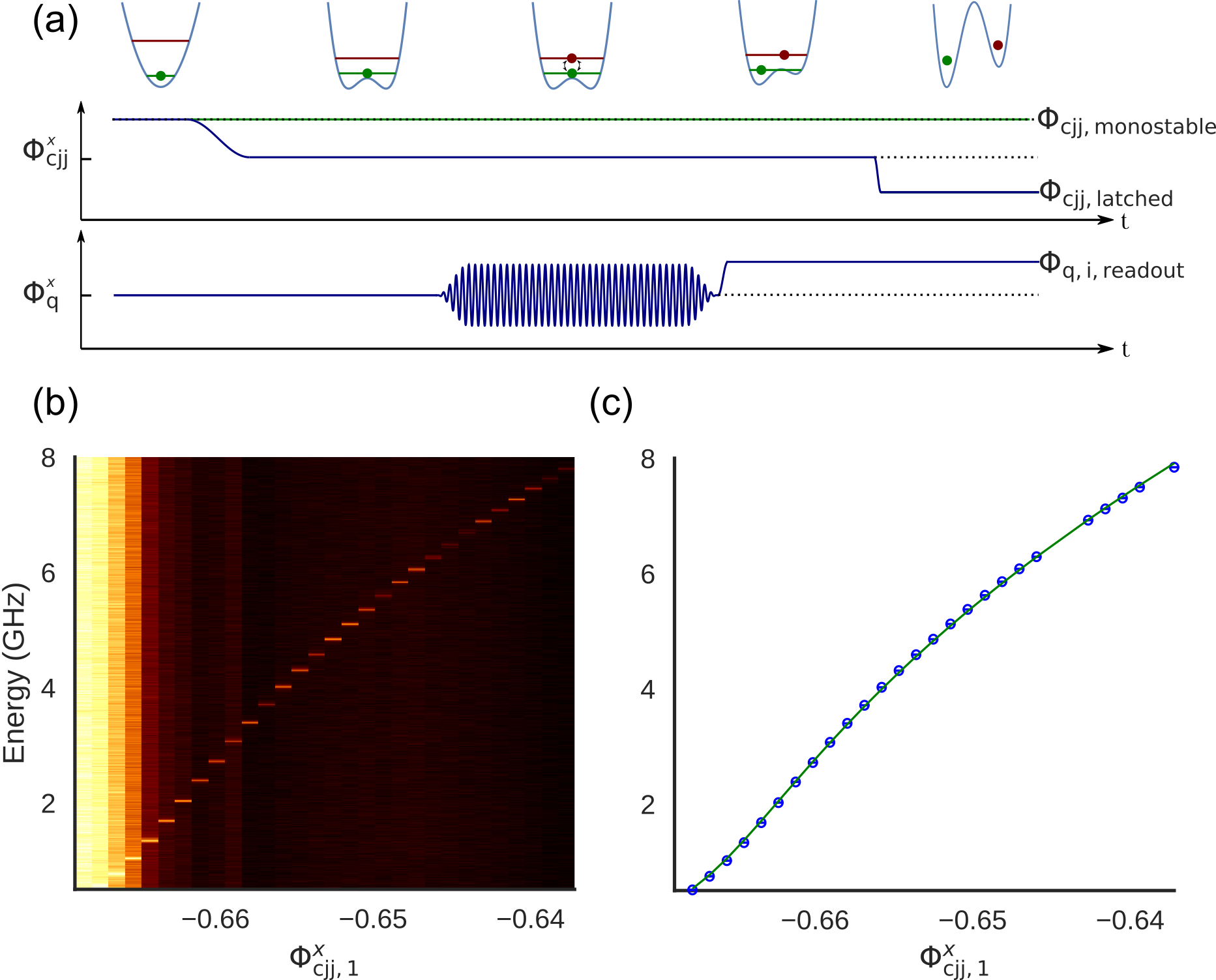}
\caption{(a) Pulse sequence and the effective qubit potential at each pulse segment for the single-qubit spectroscopy experiments. (b) Color plot of the microwave spectroscopy data versus the flux bias $\Phi_\text{cjj}^x$ that controls the barrier height for Qubit 1. (c) Extracted spectroscopy peak positions that represent the qubit frequency versus $\Phi_\text{cjj}^x$ and a fit to the numerical rf-SQUID model.}
\label{fig:1q-spectroscopy}
\end{figure}

The qubit energy eigenstates can be characterized by performing microwave spectroscopy. In Fig.~\ref{fig:1q-spectroscopy}{(a)}, we show the pulse sequence for measuring the qubit spectroscopy and the effective qubit potential at each segment of the pulse sequence. The qubit is first initialized in its ground state by adiabatically preparing a symmetric double-well potential with a relatively low tunneling barrier. In this limit, the qubit frequency, which is set by the tunneling amplitude, is much larger than $1$GHz $(\Delta \gg k_B T)$. Then, we apply a microwave pulse $\Phi_\text{q}^x$ to excite the qubit to its excited state. We sweep the frequency of the microwave pulse from 0.5GHz to 8GHz to probe all excited states in that range. After the microwave pulse, an adiabatic tilt followed by a quench, that increases the barrier height, is applied to the qubit to project the qubit ground and excited states to the clockwise and counter-clockwise persistent current states for readout. The same pulse sequence was used Ref.~\cite{Quintana17}. In this experiment, we use a rise time of 1~ns for both the adiabatic tilt and the quench on the barrier height. Throughout the single qubit spectroscopy experiments, we keep the magnetic coupling strength at $M_{12}=0$ and the other qubit at $\Phi_{\text{cjj}}^x = 0.5\Phi_0$. In Fig.~\ref{fig:1q-spectroscopy}{(b)}, we show the spectroscopy versus $\Phi_\text{cjj,1}^x$ for Qubit 1.  The extracted qubit frequency from this figure along with the fit to rf-SQUID numerical model is shown in Fig.~\ref{fig:1q-spectroscopy}{(c)}.

Effective single-qubit $\Delta$ corresponds to the energy gap between the ground state and the first excited state in the single-qubit spectroscopy measurements. Once we have an accurate model that predicts the location of the first excited state, the appropriate $\Phi_{\text{cjj},i}^x$ that needs to be applied to set the qubit to a given $\Delta$ is determined using the model extracted above. Note that the effective single-qubit $\Delta$ is influenced by the capacitive coupling that is always present. As a result the individual qubit dynamics are always affected by the presence of the secondary qubit even if it is in monostable state ($\Phi_{\text{cjj}}^x = 0.5\Phi_0$). 


\subsection{Two-qubit spectroscopy}

\begin{figure*}[!]
\includegraphics[width=140mm]{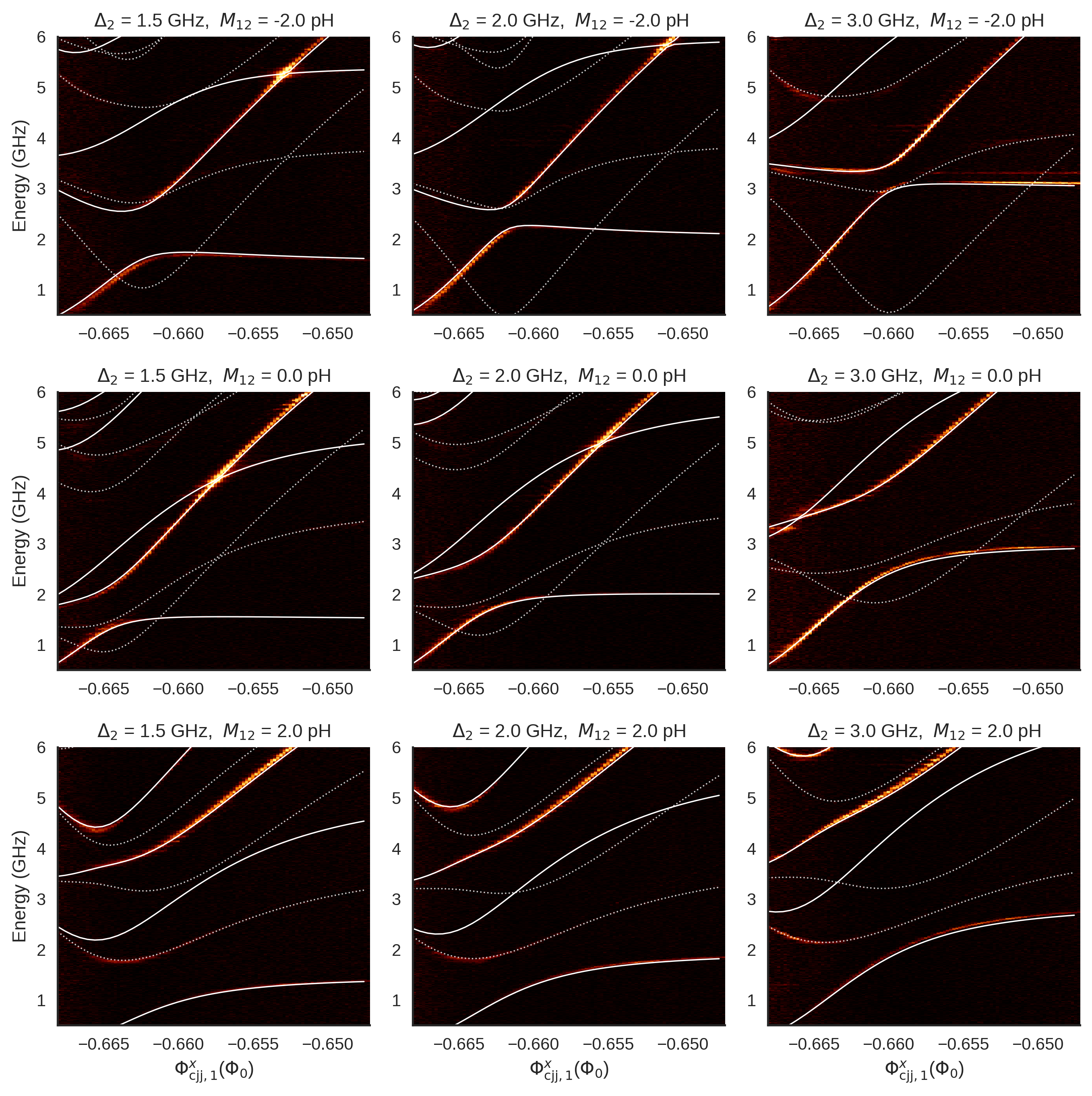}
\caption{Two-qubit spectroscopy versus $\Phi_{\text{cjj},1}^x$ for $\Delta_2/h  \approx 1.5$~GHz, 2.0~GHz, and 3.0~GHz and $M_{12} = 0$~pH and $\pm 2$~pH. The solid and dashed lines represent numerical calculations of energy differences. The solid lines correspond to excitations from the ground state and the dashed lines correspond to excitations from the first excited state.}
\label{fig:2q-spectroscopy-extra}
\end{figure*}

The coupled system energies can be probed by microwave spectroscopy in a way that is similar to single-qubit spectroscopy. In the main text we show the pulse sequence of the two-qubit spectroscopy. Although a microwave pulse is only applied to Qubit 1, the energy eigenstates of both qubits are read out by simultaneous adiabatic tilts followed by quench. We plot the spectroscopy, as the excited state population of Qubit 1, versus $\Phi_{\text{cjj},1}^x$ that controls the barrier height of Qubit 1. The barrier height of the second qubit is set to an effective single-qubit tunneling amplitude of $\Delta_2$. Using the previously extracted circuit parameters and initializing the fitting procedure by the design values for the unknown parameters, we extract $C_1 = 119.5$~fF, $C_2 = 116.4$~fF, and $C_{12} = 132.0$~fF and $L_{\text{cjj},1(2)} = 17.016(17.175)$pH. At this point, we have all the circuit parameters of this system as summarized in the following table: 
\begin{widetext}
\begin{center}
\begin{tabular}{ |c|c|c|c|c|c|c|c|c|c| } 
\hline
Qubit & $I_{c,i} (\mu\text{A})$ & $L_i (\text{pH})$ & $L_{\text{\text{cjj}},i} (\text{pH})$ & $C_i (\text{fF})$ & $C_{12} (\text{fF})$ & $|M_{12}|(\text{pH})$ \\ 
\hline
Q1 & 3.22697$\pm$4.1e-5 & 231.633$\pm$4.5e-3 & 17.02$\pm$6e-2 & 119.5$\pm$0.89 & \multirow{2}{*}{132.0$\pm$1.54} & \multirow{2}{*}{$\le 8.145$} \\
Q2 & 3.15711$\pm$3.6e-5 & 238.981$\pm$04.1e-3 & 17.17$\pm$7.9e-2 & 116.4$\pm$1.04 & & \\ 
\hline
\end{tabular}
\end{center}
\end{widetext}

We also perform similar two-qubit spectroscopy for various $\Delta_2$ and $M_{12}$ at $h_1 = h_2 = 0$ and compare the system energy spectra with numerical simulation using previously extracted parameters. In Fig.~\ref{fig:2q-spectroscopy-extra}, we show the two-qubit spectroscopy data and calculated energy spectra of the system for all the combination of $\Delta_2/h  = 1.5$~GHz, 2.0~GHz, and 3.0~GHz and $M_{12} = 0$~pH and $\pm 2$~pH. Very good agreement between the numerical model and the data is achieved. In all figures the solid (dashed) lines represent excitations from the ground (first excited) state.

\section{Pulse distortion compensation for coherent oscillations}

Short duration pulse ($\le 10$~ns) distortion imposes a great limitation on the fidelity of coherent qubit operation. In our experiment, the qubit control involves applying fast pulses to lower and rise the tunneling barrier of the flux qubits.  These pulses have short rise and fall times ($\approx 200$~ps). Here, we discuss our method of measuring and correcting pulse distortion in-situ.

\begin{figure}[h!]
\includegraphics[width=80mm]{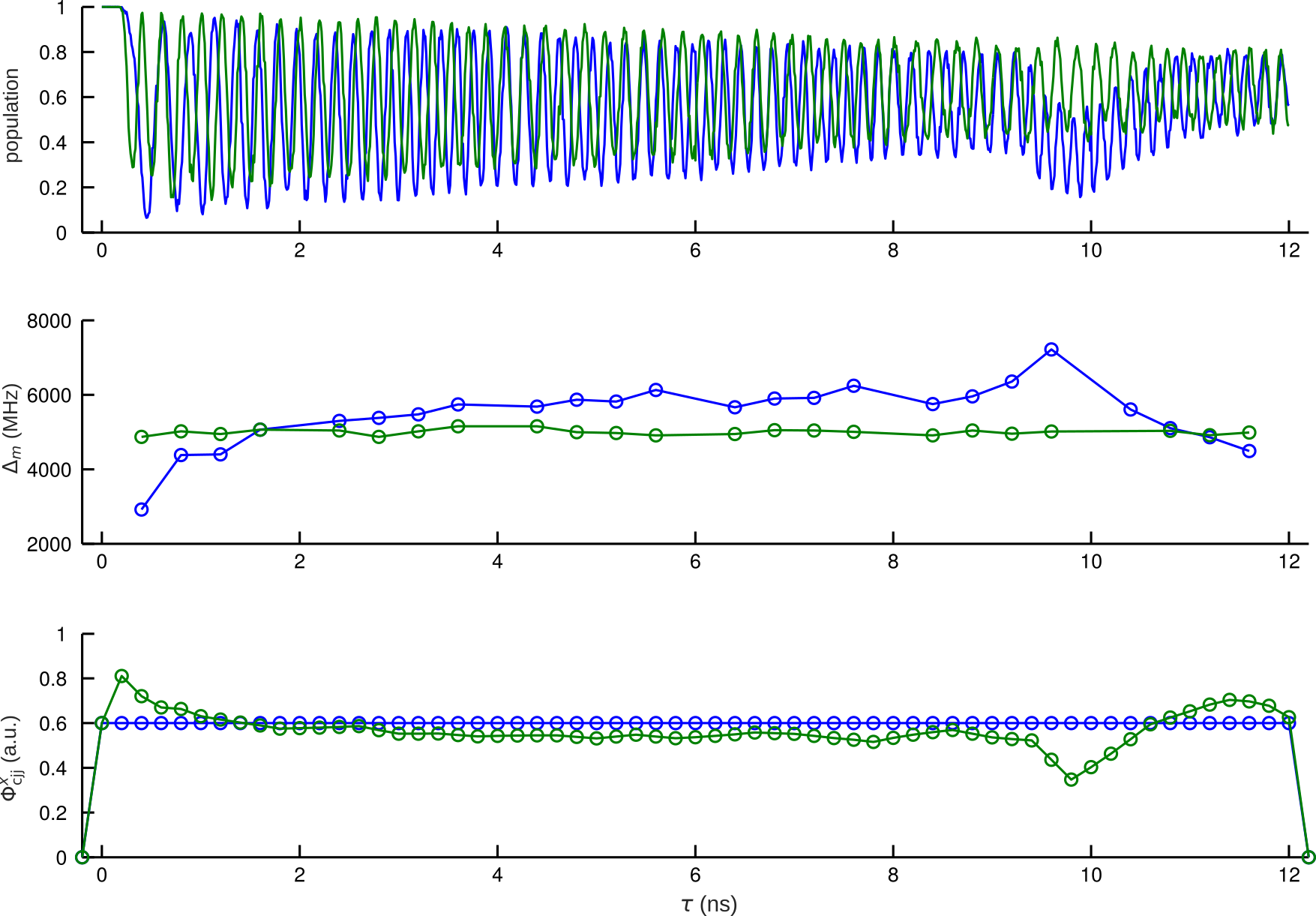}
\caption{Characterization and compensation of pulse distortion. The coherent oscillation data, measured tunneling amplitude $\Delta_m$, and applied barrier pulses without (blue) and with (green) pulse distortion corrections.}
\label{fig:pulse-compen}
\end{figure} 

We first measure the single-qubit coherent oscillations at $\Delta/h = 5$~GHz, where the qubit population in the computational basis is supposed to oscillate at the frequency $\Delta$. Pulse distortion is mainly caused by reflections. As a result, the $\Phi_{\text{cjj}}^x(t)$ signal reaching the qubit deviates from the ideal square pulse which may distort the frequency of the coherent oscillations from the target $\Delta$ at a given time $\tau$. To measure the distorted pulse in the time domain, we slice the coherent oscillation data into small time windows, typically with a length of 400~ps that contains more than one period of oscillation. We then extract the coherent oscillation frequency $\Omega(\tau)$ in each slice starting at time $\tau$ and treat this frequency as a measurement of the instantaneous $\Delta_m(\tau) = \Omega(\tau)$. The pulse distortion at any given time is then calculated as $\tau$ by $\delta\Phi_{\text{cjj}}^x(\tau) = (\Delta_m(\tau) - \Delta)/\frac{\partial \Delta}{\partial \Phi_{\text{cjj}}^x}$ where $\frac{\partial \Delta}{\partial \Phi_{\text{cjj}}^x}$ can be evaluated numerically using the qubit model. To correct for the pulse distortion, we simply apply the first order correction to the applied pulse with $\Phi_{\text{cjj,corr}}^x(\tau) = \Phi_{\text{cjj}}^x(\tau) + \delta\Phi_{\text{cjj}}^x(\tau)$ at the same time $\tau$. As the above correction at time $\tau$ may lead to distortions at times $t>\tau$, we iterate the entire measurement and correction procedure until the corrected pulse converges. In practice, this pulse distortion correction procedure converges within 5 iterations. In Fig.~\ref{fig:pulse-compen}, we show the coherent oscillation data and the extracted instantaneous tunneling amplitude $\Delta_m$ with uncorrected pulse (blue) and corrected(green) pulses $\Phi_{\text{cjj}}^x$, shown respectively. The measured tunneling amplitude $\Delta_m/h$ is fixed to the target value 5~GHz after the pulse distortion compensation is applied, which significantly increases the fidelity of the coherent oscillation.

\section{Coherent oscillation protocol}
\label{CohOsc}

\begin{figure*}[hb]
\includegraphics[width = 150 mm]{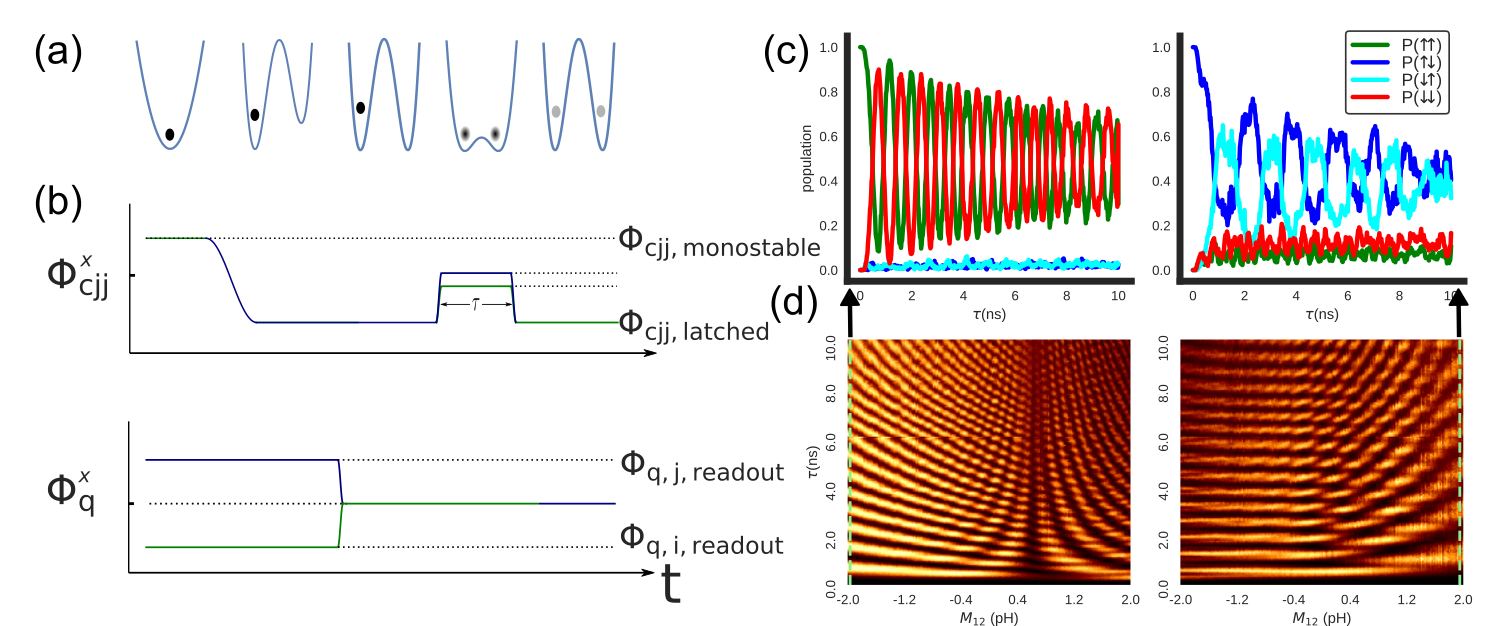}
\caption{Two-qubit coherent oscillations. (a) Pulse sequence and the effective qubit potential during each segment of the pulse. The potential landscape represents a diagonal cut across the landscape depicted in Fig.~1(a) of the main text.
\ref{fig:schematics}. 
The direction depends on the initial tilt applied on the two qubits. (b,c) Data (b) and simulations (c) of the coherent oscillations between the low-energy states versus the magnetic coupling strength $M_{12}$. For $M_{12} < 0$, the qubits are initialized in $|\!\uparrow\uparrow\rangle$ and measure the probability of the state $|\!\downarrow\downarrow\rangle$. For $M_{12} \ge 0$, the qubits are initialized in $|\!\downarrow \downarrow\rangle$ and the population of $P_{\uparrow \downarrow}$ and $P_{\uparrow \uparrow}$ are plotted respectively.(a,b) Two-qubit coherent oscillation between the computational basis states with strong FM coupling at $M_{12} = -2$~pH (a) and strong AFM coupling at $M_{12} = 2$~pH (b). (c,d) Color plots of the two-qubit coherent oscillation versus the magnetic coupling strength $M_{12}$ with the system initially prepared at $|\!\downarrow\downarrow\rangle$ (c) and $|\!\downarrow\uparrow\rangle$ (d) respectively.}
\label{fig:coherent-oscillation2}
\end{figure*}

The pulse sequence and the effective two-qubit potential at each pulse segment of a coherent oscillation protocol is shown in Fig.~\ref{fig:coherent-oscillation2}. The qubits are initialized in two steps. A large flux-bias, $\Phi_{\text{q,prep}}^x$, is applied to tilt the potential in its monostable state. Next, the tunneling strength is reduced, keeping the potential tilted. Once both qubits are prepared in a computational state controlled by the signs of $\Phi_{\text{q,prep}}$ pulses, the tilts are removed. The coherent oscillations are induced by reducing the barriers of both qubits simultaneously to allow quantum fluctuations drive the coherent dynamics of the system. As the computational basis states are not the eigenstates of the total Hamiltonian, the system undergoes coherent oscillations between the computational basis states with near-degenerate Ising energies. After some time $\tau$, the states are read out by measuring qubit-persistent currents after simultaneously quenching both qubits. The rise and fall times of the pulse, about 200~ps, are much faster than the dynamics of the qubits to snapshot the qubit population at the end of evolution. We repeat this process for a range of dwell times $\tau$ and magnetic coupling strengths. This protocol works equally well for a single-qubit if the qubit potential is kept at its monostable state.

In coherent oscillation experiments, the system is prepared and measured in one of the four computational basis states $|\!\uparrow\uparrow\rangle$, $|\!\uparrow\downarrow\rangle$, $|\!\downarrow\uparrow\rangle$, and $|\!\downarrow\downarrow\rangle$. In Fig.~\ref{fig:coherent-oscillation2}, we show that the coherent oscillations are indeed mostly between the two low-energy configurations at strong magnetic couplings. At strong FM coupling with $M_{12} = -2$~pH, the coherent oscillations occur between $|\!\uparrow\uparrow\rangle$ and $|\!\downarrow\downarrow\rangle$, whereas the coherent oscillations occur between $|\!\uparrow\downarrow\rangle$ and $|\!\downarrow\uparrow\rangle$ at strong AFM coupling with $M_{12} = 2$~pH. 

\begin{figure}[!]
\includegraphics[width=80mm]{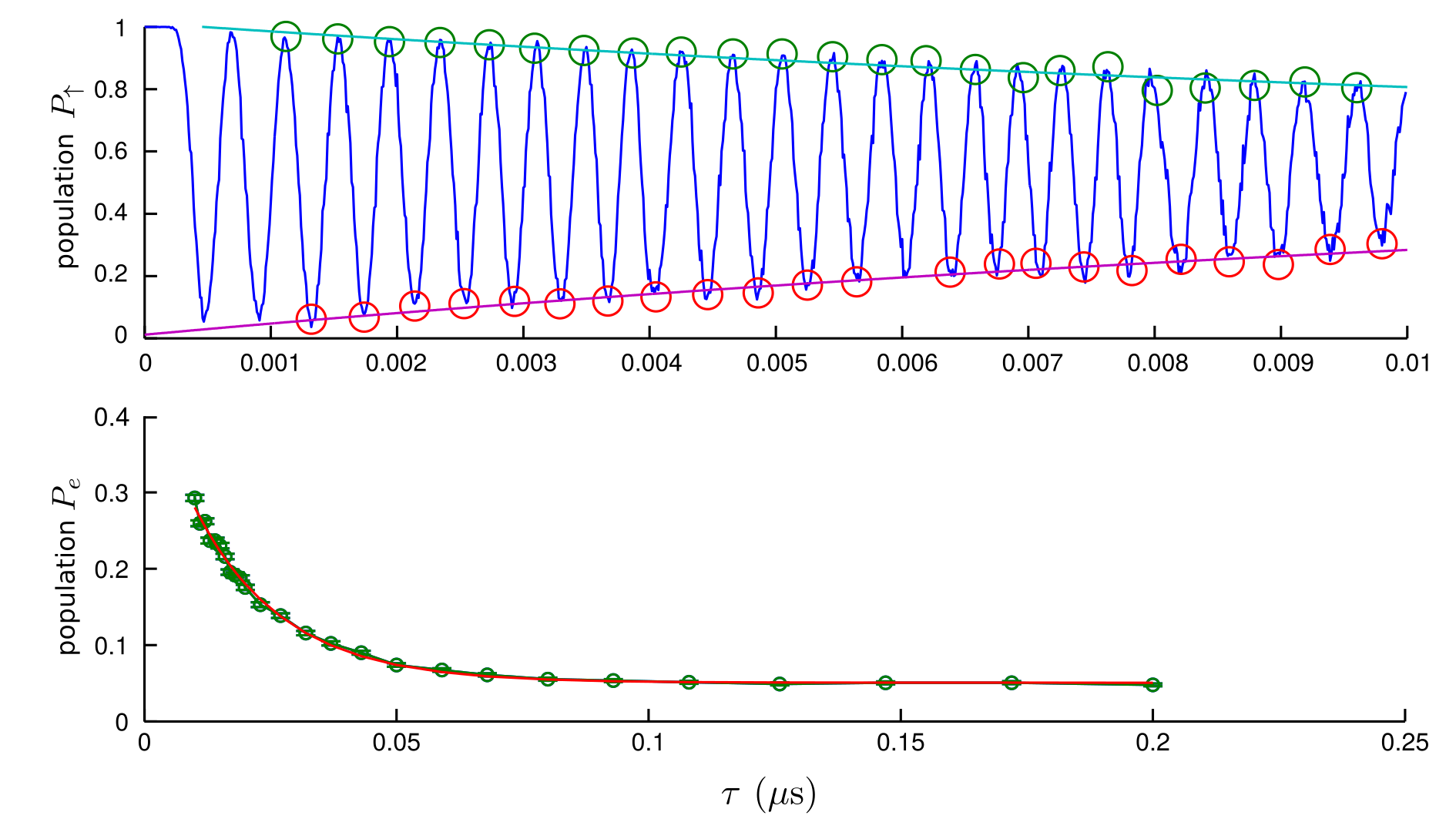}
\caption{$T_1$ and $T_2$ measurements on Qubit 1 at $\Delta = 2$~GHz and at zero bias, $h_1 = 0$. By fitting the decay envelop of the coherent oscillation to an exponential function, we obtain $T_2 = 16.2$~ns. By fitting the energy relaxation of the first excited state, we obtain $T_1 = 17.5$~ns.}
\label{fig:coherence}
\end{figure}

\section{Coherence}

The phase coherence time $T_2$ between the energy eigenstates of each qubit can be characterized by measuring single-qubit coherent oscillations. As in Eq.~(\ref{UpDown}), at zero bias we define the computational basis as $\ket{\uparrow}{=}\frac{\ket{\chi_1} {+} \ket{\chi_2}}{\sqrt{2}}$ and $\ket{\downarrow} {=} \frac{\ket{\chi_1} {-} \ket{\chi_2}}{\sqrt{2}}$ for each qubit. In Fig.~\ref{fig:coherence}, we show the coherent oscillations of Qubit 1 at $\Delta_1 = 2$~GHz. A fit to the decay envelop of the oscillations yields $T_2 = 16.2$~ns. Energy relaxation time $T_1$ at the same operation point is characterized by measuring the qubit excited state population versus the delay time between initialization and readout. By fitting the excited-state population decay to an exponential function, we extract $T_1 = 17.5$~ns. We perform the same characterization on Qubit 2 and obtain $T_2 = 14.3$~ns and $T_1 = 17.4$~ns at $\Delta_2 = 2$~GHz. We have also fabricated uncoupled qubits with the same parameters and observed the same coherence parameters, confirming that the capacitive coupling does not degrade qubit coherence.

\end{document}